\documentclass[superscriptaddress,twocolumn,showpacs,preprintnumbers,amsmath,amssymb,pra,10pt,aps]{revtex4-1}
\usepackage{graphicx}
\usepackage{dcolumn}
\usepackage{bm}
\usepackage{subfigure}
\usepackage{multirow}
\usepackage{soul}
\usepackage{float}
\usepackage[normalem]{ulem}
\usepackage[usenames,dvipsnames]{color}
\usepackage{bbold}
\usepackage{hyperref} 
\hypersetup{backref,pdfpagemode=FullScreen,colorlinks=true,linkcolor=NavyBlue,citecolor=BrickRed}

\begin{document}

\title{Slave-spin mean field for broken-symmetry states: N\'eel antiferromagnetism and its phase separation in multi-orbital Hubbard models}
\author{Matteo Crispino}
\affiliation{LPEM, ESPCI Paris, PSL Research University, CNRS, Sorbonne Universit\'e, 75005 Paris France}
\author{Maria Chatzieleftheriou}
\affiliation{LPEM, ESPCI Paris, PSL Research University, CNRS, Sorbonne Universit\'e, 75005 Paris France}
\affiliation{CPHT, CNRS, \'Ecole polytechnique, Institut Polytechnique de Paris, 91120 Palaiseau, France}
\author{Tommaso Gorni}
\altaffiliation{Present address: CINECA National Supercomputing Center, Casalecchio di Reno, I-40033 Bologna, Italy}
\affiliation{LPEM, ESPCI Paris, PSL Research University, CNRS, Sorbonne Universit\'e, 75005 Paris France}
\author{Luca de' Medici}
\affiliation{LPEM, ESPCI Paris, PSL Research University, CNRS, Sorbonne Universit\'e, 75005 Paris France}

\date{\today}

\begin{abstract}
We introduce the generalization of the Slave-Spin Mean-Field method to broken-symmetry phases. Through a variational approach we derive the single-particle energy shift in the mean-field equations which generates the appropriate self-consistent field responsible for the stabilization of the broken symmetry. With this correction the different flavours of the slave-spin mean-field are actually the same method and they give identical results to Kotliar-Ruckenstein slave-bosons and to the Gutzwiller approximation. We apply our formalism to the N\'eel antiferromagnetic state and study it in multi-orbital models as a function of the number of orbitals and Hund's coupling strength, providing phase diagrams in the interaction-doping plane. We show that the doped antiferromagnet in proximity of half-filling is typically unstable towards insulator-metal and magnetic-non magnetic phase separation. Hund's coupling extends the range of this antiferromagnet, and favors its phase separation.
\end{abstract}


\maketitle

\section{Introduction}

Antiferromagnetism (AF) is one of the hallmarks of strong electronic correlations. 
Mott insulators, their most striking manifestation, are indeed typically antiferromagnetic at half-filling and low temperature. The Cuprates provide a prominent example but many others are found among 3d transition-metal oxides, like (Ca,Sr)MnO$_3$\cite{Takeda-SrMnO3_AF} and LaCrO$_3$\cite{Bansal-LaCrO3_AF}, and pnictides like BaMn$_2$As$_2$\cite{McNally_BaMn2As2_Mott_Hund}, and also among 4d materials like Ca$_2$RuO$_4$\cite{Nakatsuji-Ca2RuO4_AF} and SrTcO$_3$\cite{Rodriguez-SrTcO3_AF}, molecular solids like Fullerenes\cite{Takabayashi-C60_Antiferro} and 2D organic superconductors\cite{Lefebvre_Jerome-k-BEDT}, etc.
 
Moreover cold-atomic systems, providing the cleanest realization of fermionic models on a lattice, do indeed show antiferromagnetic ordering\cite{Mazurenko-ColdAtoms_AF}.

In fact AF naturally arises when considering the prototypical half-filled Hubbard model in the limit of strong interactions, where it is a Mott insulator\cite{Auerbach}. There, the effective low-energy model is one of well-formed local moments interacting with short-range couplings. These have typically a super-exchange component favoring opposite spin alignment on neighboring sites, which can dominate (like e.g. in bipartite lattices with nearest-neighbor hoppings), or can be contrasted by sub-dominant (e.g. ferromagnetic) couplings or frustrating geometries.

When interactions are weak on the other hand the low-energy excitations are quasiparticles and the paramagnetic Fermi liquid is unstable towards magnetic ordering if the bare magnetic susceptibility is peaked, which typically derives from Fermi-surface nesting\cite{Moriya_book}. In the single band Hubbard model on a bipartite lattice this happens for the AF q-vector typically at half-filling in absence of next-nearest neighbor hopping. A gap opens at the Fermi level in that case, due to the doubling of the lattice unit cell, and the system is dubbed a "Slater" band insulator.

In the two mentioned extreme limits, a static mean-field treatment (Hartree-Fock at weak interaction and a Weiss mean field of the emergent Heisenberg model at strong interactions) can be satisfactory for an approximate description of these ordered phases\cite{Pavarini_Juelich_Magnetism}. At intermediate couplings however, methods capable to describe the gradual localization of the electrons are needed to bridge between these two pictures.

Moreover at non-integer fillings charge fluctuations disrupt the Mott insulating behaviour and this causes the simple Heisenberg low-energy picture to lose its validity. Hence, the description of the crossover between this metallic phase and the metal arising at weak coupling from the doping of the Slater insulator is even more subtle.
These crossovers are outside the reach of the respective (weak-coupling and strong-coupling) perturbative methods.
On the other hand exact numerical methods typically struggle with the size of the system they can describe and thus with the representativity of their results for the thermodynamic limit.
A streamline of methods can instead be formulated directly in this limit if the range of correlations is restricted (typically to just on-site ones). Among these the Gutzwiller Approximation\cite{Gutzwiller} (GA) and the Slave-Boson formulation (which in the mean-field approximation - SBMF - is equivalent to the former\cite{kotliar_ruckenstein,fresard}) have reached some success. They have preceded the more elaborate and now widespread Dynamical Mean-Field Theory (DMFT)\cite{georges_RMP_dmft}, which provides a local but fully frequency-dependent electronic self-energy, of which the former methods give just a low-frequency approximation. More recently the Slave-Spin Mean Field\cite{demedici_Slave-spins,Hassan_CSSMF,YuSi_LDA-SlaveSpins_LaFeAsO} (SSMF), very similar in spirit to SBMF, has gained attention because of its agility, especially in multi-orbital cases and for realistic simulations of materials.

Indeed, among the aforementioned physical examples, only the Cuprates, the 2D organics and cold atoms in a suitably designed optical lattice are single-band systems when they become metallic. In fact, most correlated materials are multi-orbital.

In these systems the orbital and spin degrees of freedom produce a larger variety of local configurations, possibly giving rise to an equal variety of long-range ordered phases. Moreover, all integer fillings give rise to Mott insulators at strong interaction strengths. 
The focus of this article is on N\'eel antiferromagnetism at and near half-filling.

The work we report is twofold. First we present the generalization of the Slave-spin method introduced in Refs. \cite{demedici_Slave-spins,Hassan_CSSMF} to broken-symmetry phases (section \ref{sec:method}). We also discuss previous attempts to this aim and variants of the method. We demonstrate that, at zero temperature, the mean fields of the slave-spin and slave-bosons methods coincide in both magnetic and non-magnetic phases, and thus all coincide with the Gutwziller approximation.
Second, we apply the new method to the study of the degenerate multi-orbital Hubbard model at, and around, half-filling. We study in detail the 2-orbital and 3-orbital cases (section \ref{sec:2_3orb}) as a function of the local interaction strength U and for different relative strenghts ($J/U$) of the onsite Hund's coupling $J$. Our method portrays correctly the progressive localisation of the electrons as a function of U and the expected formation of high-spin insulating AF states at strong interactions. We show that a larger Hund's coupling favors the formation of large localized moments. Interestingly, however, in the 3-orbital model a low-spin to high-spin first-order transition happens as a function of U for J=0. 

A key finding of our study is that the metallic phase obtained by doping the antiferromagnet is often unstable towards phase separation. The stable mixture which is formed is determined through Maxwell constructions and can be, depending on the specific situation, a coexistence of 1) an AF insulator and a paramagnetic (PM) metal, 2) an AF insulator and an AF metal, 3) two metals, one AF and one PM.

In Sec.~\ref{sec:corr} we illustrate the method's capability to describe correctly the itinerant-to-localized magnetism crossover in the different models we explore, and the non-trivial dependence of the correlation strength on doping and interaction strength.

Finally, in Sec.~\ref{sec:BaCr2As2}, we showcase the potential of the method in the context of realistic \emph{ab-initio} simulations of materials through a comparison of Density-Functional Theory (DFT)+SSMF and DFT+DMFT for the case of BaCr$_2$As$_2$. The agreement is excellent.
The appendix shows further benchmarks with DMFT, validating our method.

\section{Model and method}\label{sec:method}

In this paper, we consider the multi-orbital Hubbard model, whose
Hamiltonian reads:

\begin{align}
\label{eqn:Hubbard}
\hat{H}-\mu\hat{N} & ={\sum_{ij,m m',\sigma}}t_{ij}^{mm'}\hat{d}_{im\sigma}^\dagger\hat{d}_{jm'\sigma}\nonumber \\
 & +\sum_{im\sigma}\left(\epsilon_{m}-\mu\right)\hat{n}_{im\sigma}+\hat{H}_{\mathrm{int}},
\end{align}
where $\hat{d}_{im\sigma}^{\dagger}$
($\hat{d}_{im\sigma}$) is the creation (annihilation) operator for
an electron at site $i$, in orbital $m$, with spin $\sigma$, $\hat{n}_{im\sigma}=\hat{d}_{im\sigma}^{\dagger}\hat{d}_{im\sigma}$
is the number operator, and $\hat{N}=\sum_{im\sigma}\hat{n}_{im\sigma}$. 
The $t_{ij}^{mm'}$ are real and describe the hoppings between orbitals $m$ on site i and orbital
$m^\prime$ on site $j$; 
$\epsilon_{m}$ is  the on-site energy in orbital
$m$ (we take $t_{ii}^{mm'}=0$) and $\mu$ is the chemical potential.  

$\hat{H}_{\mathrm{int}}$ is the local, multi-orbital interaction
Hamiltonian, that we take customarily in the density-density
form (i.e. we drop spin-flip and pair-hopping terms):
\begin{align}
\label{eqn:Interaction}
\hat{H}_{\mathrm{int}} & =U\underset{i,m}{\sum}\hat{n}_{im\uparrow}\hat{n}_{im\downarrow}+U'\underset{i,m\neq m'}{\sum}\hat{n}_{im\uparrow}\hat{n}_{im'\downarrow}\nonumber \\
 & +(U'-J)\underset{i,m<m',\sigma}{\sum}\hat{n}_{im\sigma}\hat{n}_{im'\sigma},\end{align}
Here $U$ and $U'$ are the Coulomb repulsion on the same
orbital and on different ones respectively, and $J$ is the local exchange, i.e. Hund's
coupling. We assume \cite{Georges_annrev} $U'=U-2J$ throughout the paper.

\subsection{Slave spins and broken-symmetry phases}

The model is solved in the slave-spin mean field\cite{demedici_Slave-spins,Hassan_CSSMF} (see \cite{demedici_Vietri} for a pedagogical introduction - this method is sometimes termed Z$_2$-SSMF to distinguish it from its variant U(1)-SSMF\cite{YuSi_LDA-SlaveSpins_LaFeAsO}, based on the symmetry group of the representation). 

SSMF has up to now been applied only to phases without a spontaneous broken symmetry. Only sparse attempts at cluster generalizations (that is, beyond the simplest local mean-field, including short-range correlations) have been found able to address charge-\cite{Hassan_CSSMF} or spin-ordered\cite{Wei-Cheng_Lee-Cluster_SS_AF} phases\footnote{Note however that Antiferromagnetism was addressed in \cite{Ruegg-Z2-Slave-Spin} in a less general Z2-slave-spin scheme, only for the case of the single-band Hubbard model, with a different mean-field decoupling and in presence of fluctuations beyond the mean field.}. 

It is instead important to investigate broken-symmetry phases in the purely local-correlation framework for several reasons. First, in order to disentangle the effect of local vs short-range correlations. Second, to have an unambiguous method, free from the issue of the multiple sizes and shapes of the clusters, which have to be explored to reach an unbiased result. Third, and most importantly, because the computational power can be used to tackle multi-orbital, and thus realistic systems (possibly in conjunction with electronic structure calculations). 

An analysis of the reason why SSMF seemed unable to access symmetry breaking was performed in Ref.~\onlinecite{Georgescu_Slave-Spin_Magn}. It was found that the on-site energy shifts due to correlations, acting in this framework as a Weiss-like effective field, were unable to sustain self-consistently the spontaneous symmetry breaking allowed \emph{a priori} by the method. An external field was added there, to artificially explore the energy landscapes and search for minima with broken symmetry \cite{Georgescu_Slave-Spin_Magn,Georgescu-BOSS}.

Here we show instead, using a variational derivation of the Z$_2$-SSMF mean-field equations, that these were incomplete, since a term providing precisely a local field was missing in the original formulation based on simple mean-field decoupling of the Hamiltonian.

Incidentally this term, for real-valued self-consistent Hamiltonians, coincides to the one present in the U(1) variant, rendering the two mean fields essentially identical \footnote {This obviously implies that the U(1) SSMF should be able to investigate symmetry breaking as is, in our understanding. This was however not done to date, to the best of our knowledge}.

We furthermore show that, including the proper energy shift derived here, the SSMF method yields exactly the same results as SBMF everywhere we could check (both for broken- and unbroken-symmetry phases). This means that most probably SSMF, SBMF and GA are strictly equivalent methods for the ground state, i.e. at equilibrium for T=0.

\subsection{Slave-spin representation}

In the slave-spin representation, to every local electronic one-particle state (created by $\hat{d}_{im\sigma}^{\dagger}$) are associated, in a dual, larger Hilbert space, both a fermionic state (created by $\hat{f}_{im\sigma}^{\dagger}$) and a spin-1/2 variable, of z-component $S_{im\sigma}^{z}$. They both carry the same indices of the original d fermion, and the slave-spin has no special relation - and should not be confused - with the physical spin. An occupied local state $|1\rangle_d$ in the original space is then associated, in the dual Hilbert space, to the product of the occupied state for the corresponding f fermion and to the "up" state of the associated "slave"-spin $|1\rangle_f|\uparrow\rangle_s$. Likewise an empty d-state $|0\rangle_d$ is associated to the empty fermionic state and "down" state of the slave spin $|0\rangle_f|\downarrow\rangle_s$.
These "physical" states satisfy the operatorial relation
\begin{equation}
\label{eqn:constraint}
\hat{f}_{im\sigma}^{\dagger}\hat{f}_{im\sigma}=\hat{S}_{im\sigma}^{z}+\frac{1}{2},
\end{equation}
which can be used to distinguish them from the remaining combinations  $|1\rangle_f|\downarrow\rangle_s$ and $|0\rangle_f|\uparrow\rangle_s$ which have no physical counterpart. 

This is apparently a complication, but it turns out that approximations performed on the larger space can be less severe than if applied directly to the original system.

The operators of the original space are mapped on operators having the same action on the physical states of the larger space, while their action on the unphysical states can be chosen freely, in principle. In practice this freedom is used to gauge approximated treatments e.g. to reproduce known limits and possibly yield the most physical results.

Hence the number operator $\hat{n}_{im\sigma}=\hat{d}_{im\sigma}^\dagger\hat{d}_{im\sigma}$ can be equivalently mapped in the number operator for the pseudo-fermions $\hat{n}_{im\sigma}^{f}=\hat{f}_{im\sigma}^\dagger\hat{f}_{im\sigma}$ or in $\hat{S}_{im\sigma}^{z}+\frac{1}{2}$, which, because of eq.(\ref{eqn:constraint}), gives the same result by construction on the physical states.
The off-diagonal destruction (creation) operator $\hat{d}_{im\sigma}^{(\dagger)}$ can be expressed by $\hat{f}_{im\sigma}^{(\dagger)}\hat{O}_{im\sigma}^{(\dagger)}$.
The general form of the $\hat{O}_{im\sigma}$ operator \cite{Hassan_CSSMF} is $\hat{O}_{im\sigma}=\hat{S}_{im\sigma}^{-}+c_{im\sigma}\hat{S}_{im\sigma}^{+}$, where $c_{im\sigma}$ is an arbitrary gauge embodying the aforementioned freedom. Throughout this paper, we choose  $c_{im\sigma}$ to be a given function of the average values of a generic fermionic operator $\hat{F}_{im\sigma}$ (that we will specify later).
The interaction Hamiltonian, which includes only density-density terms, can be written in terms of the $\hat{S}^{z}$ operators only. We will refer it as $\hat{H}_{\mathrm{int}}[\hat{S}^{z}]$.

\subsection{Variational approach to the slave-spin Hamiltonian}

We here derive a mean-field approximation at T=0 using the variational principle and an ansatz on the ground state $|\Psi_{\mathrm{tot}}\rangle$, for which we posit a factorized form i) between the fermion and slave-spin part, and ii) between each site in the slave-spin lattice. 
Point i) reads $\left|\Psi_{\mathrm{tot}}\right\rangle =\left|\Psi_{f}\right\rangle \left|\Phi_{s}\right\rangle$ and the evaluation of the 
average value of the Hamiltonian in Eq.~(\ref{eqn:Hubbard}), rewritten in the slave-spin representation, then gives:
\begin{align}
\label{eqn:energy}
\langle\hat{H}-\mu\hat{N}\rangle_{\Psi_{\mathrm{tot}}} & \!\!\!\! = \!\!\!\!\! \sum_{ij,mm',\sigma}\!\!\! t_{ij}^{mm'}\langle \hat{f}_{im\sigma}^{\dagger}\hat{f}_{jm'\sigma}\rangle _{f}\langle \hat{O}_{im\sigma}^\dagger\hat{O}_{jm'\sigma}\rangle _{s}\nonumber \\
 & +\sum_{im\sigma}(\epsilon_{m}-\mu)\langle{\Phi_{s}}|\Phi_{s}\rangle\langle \hat{n}_{im\sigma}^{f}\rangle _{f} \nonumber \\
&+\langle \hat{H}_{\mathrm{int}}[\hat{S}^{z}]\rangle.
\end{align}
 By notation, the subscripts $f$ and $s$ indicate that the average is performed with respect to the fermionic and the spin wave function respectively, because the operators act either on $|\Psi_{f}\rangle$ or $|\Phi_{s}\rangle$. 
 
We then introduce the energy functional to be minimized:
\begin{align}
\label{eqn:functional}
\mathcal{E}[|\Psi_{f}\rangle,|\Phi_{s}\rangle,\{\lambda_{im\sigma}\},{E}_{f},{E}_{s}]=\langle\hat{H}-\mu\hat{N}\rangle_{\Psi_{\mathrm{tot}}} \nonumber \\
 +\sum_{im\sigma}\lambda_{im\sigma}(\langle \hat{S}_{im\sigma}^{z}\rangle _{s}+\frac{1}{2}-\langle \hat{n}_{im\sigma}^{f}\rangle _{f})\nonumber \\
  -E_{f}[\langle \Psi_{f}|\Psi_{f}\rangle -1]-E_{s}[\langle \Phi_{s}|\Phi_{s}\rangle -1],
\end{align}
where the $\lambda_{im\sigma}$ are Lagrange multipliers enforcing (on average, due to the factorization of $|\Psi_{\mathrm{tot}}\rangle$) the slave-spin constraint of Eq.~\eqref{eqn:constraint}; the last two terms guarantee the normalization of the wave functions, via the Lagrange multipliers $E_{f}$ and $E_{s}$.
In Eq.~\eqref{eqn:functional}, $\mathcal{E}$ is a functional of the variational parameters $|\Psi_{f}\rangle$, $|\Phi_{s}\rangle$, the $\{\lambda_{im\sigma}$\}, $E_f$ and $E_s$; in its minimum, where the constraints are satisfied, it coincides with Eq.~\eqref{eqn:energy}, i.e. the total energy of the system, in its (approximated) ground state.
Proceeding with the minimization at this point would lead to two separate problems on a lattice, one of non-interacting fermions and one of interacting spins (both locally and non-locally, since the hopping terms give rise to inter-site interactions for the slave-spins).

In order to deal with more manageable calculations a further approximation is needed. The simplest one\footnote{At this stage one can proceed alternatively, e.g. to cluster approximations.} is ii), i.e. to factorize the slave-spin state as the product of single states, namely $\left|\Phi_{s}\right\rangle =\prod_i\left|\phi^{i}_{s}\right\rangle $, which is formally equivalent to a single-site (Weiss) mean-field approximation. 
It has to be noted, however, that such a single-site problem is still a many-body one, due to the onsite interactions between the slave spins. 
As a consequence of this mean field, $\langle \hat{O}_{im\sigma}^{\dagger}\hat{O}_{jm'\sigma}\rangle _{s}=\langle \hat{O}_{im\sigma}^{\dagger}\rangle _{s}\langle \hat{O}_{jm'\sigma}\rangle _{s}$ for $j\neq i$.

We now minimize the energy functional: by deriving with respect to $\lambda_{im\sigma}$ we recover the average of the constraint eq.~\eqref{eqn:constraint}, 
\begin{equation}
\label{eqn:constraint_avg}
\langle \hat{f}_{im\sigma}^{\dagger}\hat{f}_{im\sigma}\rangle=\langle \hat{S}_{im\sigma}^{z}\rangle+\frac{1}{2},
\end{equation}
and by deriving with respect to $E_f$ and the $E_s$ we obviously find the normalization conditions. By deriving with respect to $\langle \Psi_{f}|$ and $\langle \Phi_{s}|$, we are left with the following two eigenvalue problems. 

From $\frac{\delta \mathcal{E}}{\delta \langle \Phi_{s}|}=0$
we get $\hat{H}_{s}|\Phi_{s}\rangle =E_{s}|\Phi_{s}\rangle,$
with

\begin{align}
\label{eqn:Spin_Ham}
\hat{H}_{s}\!=\!\sum_i\hat{H}_{s}^i &=\!\sum_{im\sigma}(h_{im\sigma}\hat{O}_{im\sigma}^{\dagger}+\mathrm{h.c.})\nonumber\\
&+\sum_{im\sigma}\lambda_{im\sigma}\hat{S}_{im\sigma}^{z}+\hat{H}_{\mathrm{int}}[\hat{S}^{z}],
\end{align}
which is in fact simply the sum of the site-dependent, but independent from one-another, eigenvalue problems $\hat{H}_{s}^i|\phi_{s}^i\rangle =E_{s}^i|\phi_{s}^i\rangle$, where $E_s=\sum_i E_{s}^i$.
Here we have defined $h_{im\sigma}=\sum_{jm'}t_{ij}^{mm'}\langle \hat{O}_{jm'\sigma}\rangle _{s}\langle \hat{f}_{im\sigma}^{\dagger}\hat{f}_{jm'\sigma}\rangle _{f}$. 

From $\frac{\delta \mathcal{E}}{\delta \left\langle \Psi_{f}\right|}=0$
we obtain $\hat{H}_{f}|\Psi_{f}\rangle =E_{f}|\Psi_{f}\rangle $,
with: 

\begin{align}
\label{eqn:Fermionic_Ham}
\hat{H}_{f}= & \sum_{ij,mm',\sigma}t_{ij}^{mm'}\langle \hat{O}_{im\sigma}^{\dagger}\rangle_s\langle \hat{O}_{jm'\sigma}\rangle_s \hat{f}_{im\sigma}^{\dagger}\hat{f}_{jm'\sigma}\nonumber \\
 & +\sum_{im\sigma}(h_{im\sigma}\langle \hat{S}_{im\sigma}^{-}\rangle_s\frac{\partial c_{im\sigma}}{\partial\langle \hat{F}_{im\sigma}\rangle_{f}}\hat{F}_{im\sigma}+\mathrm{h.c.})\nonumber \\
 & +\sum_{im\sigma}(\epsilon_{im\sigma}-\mu-\lambda_{im\sigma})\hat{n}_{im\sigma}^{f}
\end{align}

The terms coming from the functional derivatives of the norms $\langle{\Phi_{s}}|\Phi_{s}\rangle$ and $\langle \Psi_{f}|\Psi_{f}\rangle$ appearing in Eq. (\ref{eqn:energy}) are simply absorbed into the numerical values of ${E}_{f}$ and ${E}_{s}$.

Up to this point, we have derived the variational equations under the only assumption that the gauge $c_{im\sigma}$ is a function of the average value of an operator. 
It is crucial to remark here that this posit alone makes our total Hamiltonian in the enlarged Hilbert space a self-consistent one, even before any approximation is performed. When applying the variational principle by taking the functional derivative of the energy, besides the terms coming from the dependence of the total integral on the wavefunctions used in the bra and the ket, an extra term appears (the second line in eq. \ref{eqn:Fermionic_Ham}) which comes from the dependence of the Hamiltonian itself on these wavefunctions.

Now we determine the gauge explicitly, and we do this by requiring that the known non-interacting limit (where all the hoppings in Eq.~\eqref{eqn:Fermionic_Ham} are unrenormalized, i.e. $\langle \hat{O}_{im\sigma}^{\dagger}\rangle _{s}\langle \hat{O}_{jm'\sigma}\rangle _{s}=1$) is correctly recovered in our approximate treatment, when $H_{int}=0$. In this limit the spin Hamiltonian in Eq.~\eqref{eqn:Spin_Ham} is diagonalizable analytically (see e.g. the Appendix of \cite{Hassan_CSSMF} for the derivation) since the operators with different orbitals and spin indices decouple. The value for each gauge ensuring this depends only on the corresponding fermionic density (i.e. $\hat{F}_{im\sigma}=\hat{n}_{im\sigma}^{f}$) and is:
\begin{equation}
\label{eqn:gauge}
c_{im\sigma}=\frac{1}{\sqrt{\langle \hat{n}_{im\sigma}^{f}\rangle_f (1-\langle \hat{n}_{im\sigma}^{f}\rangle_f )}}-1.
\end{equation}
Consequentially, the gauge derivative in Eq.~\eqref{eqn:Fermionic_Ham} is $\partial c_{im\sigma}/\partial\langle \hat{n}_{im\sigma}^{f}\rangle _{f}=2\eta_{im\sigma}(c_{im\sigma}+1)$ where we have defined $\eta_{im\sigma}=\frac{2\langle \hat{n}_{im\sigma}^{f}\rangle _{f}-1}{4\langle \hat{n}_{im\sigma}^{f}\rangle _{f}(1-\langle \hat{n}_{im\sigma}^{f}\rangle _{f})}$. 

The fermionic Hamiltonian thus is:
\begin{align}
\label{eqn:Final_Fermionic_Ham}
\hat{H}_{f}= & \sum_{ij,mm',\sigma} t_{ij}^{mm'}\langle \hat{O}_{im\sigma}^{\dagger}\rangle _{s}\langle \hat{O}_{jm'\sigma}\rangle _{s}\hat{f}_{im\sigma}^{\dagger}\hat{f}_{jm'\sigma}\nonumber \\
 & +\sum_{im\sigma}(\epsilon_{im\sigma}-\mu-\lambda_{im\sigma}+\lambda_{im\sigma}^{0})\hat{n}_{im\sigma}^{f}
\end{align}
where 
\begin{equation}
\label{eqn:lambda0_generic}
\lambda_{im\sigma}^{0}=2(h_{im\sigma}\langle \hat{S}^-_{im\sigma}\rangle _{s}+\mathrm{c.c.})\eta_{im\sigma}(c_{im\sigma}+1).
\end{equation}

Moreover, in the phases studied in this work one finds a real self-consistent spin Hamiltonian, i.e. $h_{im\sigma}\in \mathbb{R}$, and real eigenvectors. This gives $\langle \hat{S}^-_{im\sigma}\rangle _{s}=\langle \hat{S}^+_{im\sigma}\rangle _{s} \in \mathbb{R}$ and $\langle \hat{O}_{im\sigma}^{\dagger}\rangle _{s}=\langle \hat{O}_{im\sigma}\rangle _{s}=(c_{im\sigma}+1)\langle \hat{S}^-_{im\sigma}\rangle _{s} \in \mathbb{R}$, leading the more compact expression: 
\begin{equation}
\label{eqn:lambda0}
\lambda_{im\sigma}^{0}=4h_{im\sigma}\langle \hat{O}_{im\sigma}\rangle _{s}\eta_{im\sigma}.
\end{equation}
Incidentally, in the non-interacting limit $\langle \hat{O}_{im\sigma}\rangle _{s}=1$, this expression coincides with that of the Lagrange multiplier $\lambda_{im\sigma}=4h_{im\sigma}\eta_{im\sigma}$, which can also be calculated in a closed form in this limit\cite{Hassan_CSSMF}. Thus $\lambda_{im\sigma}$ and $\lambda_{im\sigma}^{0}$ cancel in eq. (\ref{eqn:Final_Fermionic_Ham}), which is also necessary to obtain the correct chemical potential in this limit. Notice however that the only position we made is the dependence of the gauge Eq.~\eqref{eqn:gauge} on the occupancy\footnote{The same result can be obtained if we write the occupancy in the gauge $c_{im\sigma}$ through the slave-spin equivalent $\langle \hat{S}_{im\sigma}^{z}\rangle+\frac{1}{2}$ as ensured by the constraint eq. (\ref{eqn:constraint_avg}). This will give a contribution to $H_s$ including the orbital-resolved self-consistent field conjugated to $\hat{S}_{im\sigma}^{z}$ analogous of the $\lambda_{im\sigma}^{0}$ introduced here in $H_f$. This will lead to a simple shift in the converged values of $\lambda_{im\sigma}$, but the physical results will remain unchanged.}.

All in all we obtain two Hamiltonian secular problems, with Hamiltonians Eq.~\eqref{eqn:Spin_Ham} and Eq.~\eqref{eqn:Final_Fermionic_Ham}, where the couplings in each one depend on the solutions of the other, and which coincide with the known formalism of Z$_2$-SSMF, with the exception of  the value of $\lambda_{im\sigma}^{0}$. Eq.~\eqref{eqn:lambda0_generic} was indeed absent in previous works and $\lambda_{im\sigma}^{0}$ was fixed to the numerical value of $\lambda_{im\sigma}$ determined at $U=J=0$, in order to guarantee this limit. It was however kept fixed at all other values of U and J (see e.g. the discussion in the Appendix A of Ref. \onlinecite{Chatzieleftheriou_RotSym}). On the contrary, within the present variational approach, the value of $\lambda_{im\sigma}^{0}$  descends from the functional minimisation and is thus valid for the fully interacting Hamiltonian at any $U$ and $J$. 
Operatively, the equations are solved by iterations, and the values of $c_{im\sigma}$ and $\lambda_{im\sigma}^{0}$ are thus updated at each step. 

We remark also that Eq.~\eqref{eqn:lambda0} coincides with the analogous energy shift in the U(1)-SSMF\cite{YuSi_LDA-SlaveSpins_LaFeAsO,Yu_Si-OSM_hyb_SSMF2017}, marked there as $\tilde\mu_\alpha$. The equations of the two mean-fields are thus identical (see also Ref.~\onlinecite{Pizarro_thesis}) and so are obviously expected to be the final results.

\subsection{Broken-symmetry phases}

The outlined method is general enough to study both uniform and symmetry-broken normal Fermi-liquid phases.
As in any mean-field theory the allowed symmetry breaking is to be chosen a priori, and the actual ground state of the system is determined eventually by comparing the energy of the respective stable solutions with different symmetry. 

When allowing for translational symmetry breaking we define a new, larger unit cell of the lattice (usually referred to as \emph{supercell}), encompassing the representative sites made unequal by the broken symmetry, and the translation of which tiles the lattice completely.
The single-site slave-spin wavefunctions will be allowed to differ from site to site within the supercell, i.e. a separate single-site slave-spin problem will be solved for each of the representative sites. Translational invariance will instead be assumed from supercell to supercell. 

Eqs. \eqref{eqn:Spin_Ham} and \eqref{eqn:Final_Fermionic_Ham} still hold (as long as no on-site hybridization between the orbitals is present), now with the understanding that the spacial indices run over the supercells, and the orbital index is now accompanied by another index ($\nu$, in the following) labelling the atoms within the supercell.

\section{Results on N\'eel Antiferromagnetic phases}\label{sec:2_3orb}

We here focus on N\'eel antiferromagnetic phases, where a net magnetic polarization is realized on each site, in a staggered fashion from site to site. We thus consider a generic bipartite
lattice, whose sublattices we name $A$ and $B$, i.e. we consider a unit cell with two sites, one belonging to each sublattice. 

To diagonalize the non-interacting fermionic Hamiltonian we define the Fourier transform of the creation and destruction operators for each inequivalent site $\nu=A,B$, namely $\hat{f}_{\boldsymbol{k}\nu m\sigma}^{\dagger}=\frac{1}{\sqrt{\mathcal{N}}}\underset{i}{\sum}e^{i\boldsymbol{k}\cdot\boldsymbol{R}_{i}}\hat{f}_{i\nu m\sigma}^{\dagger}$, 
with $\mathcal{N}$ to indicate the number of unit cells of the lattice and $\boldsymbol{R}_{i}$ to denote the position of the $i$-th unit cell (the relative positions of the inequivalent sites within the unit cell are immaterial at this stage, as they can be reabsorbed in the definition of the $\hat{f}_{\boldsymbol{k}\nu m\sigma}^{\dagger}$'s). The wave-vector $\boldsymbol{k}$ spans the "magnetic" Brillouin zone (MBZ) which is half the size of the original, "non magnetic" one (NMBZ). Consequently $\hat{f}_{i\nu m\sigma}^{\dagger}=\frac{1}{\sqrt{\mathcal{N}}}\underset{\boldsymbol{k}\in\mathrm{MBZ}}{\sum}e^{-i\boldsymbol{k}\cdot\boldsymbol{R}_{i}}\hat{f}_{\boldsymbol{k}\nu m\sigma}^{\dagger}$.

Then, Eq.~\eqref{eqn:Final_Fermionic_Ham} gives:
\begin{align}\label{eq:H_f_AF}
\hat{H}_{f} \! =&\!\!\!\!\underset{\boldsymbol{k}mm'\sigma}{\sum}\sqrt{Z^{A}_{m\sigma}Z^{B}_{m'\sigma}}\varepsilon_{\boldsymbol{k}}^{mm'}(\hat{f}_{\boldsymbol{k}Am\sigma}^{\dagger}\hat{f}_{\boldsymbol{k}Bm'\sigma}+\mathrm{h.c.})\nonumber \\
 +\!\underset{\boldsymbol{k}m\sigma}{\sum}&(\epsilon_{m}\!-\!\mu\!-\!\tilde{\lambda}^A_{m\sigma}\!+\!\sqrt{Z^{A}_{m\sigma}Z^{A}_{m'\sigma}}\gamma_{\boldsymbol{k}}^{mm'}\!)\hat{f}_{\boldsymbol{k}Am\sigma}^{\dagger}\hat{f}_{\boldsymbol{k}Am\sigma}\nonumber \\
 +\!\underset{\boldsymbol{k}m\sigma}{\sum}&(\epsilon_{m}\!-\!\mu\!-\!\tilde{\lambda}^B_{m\sigma}\!+\!\sqrt{Z^{B}_{m\sigma}Z^{B}_{m'\sigma}}\gamma_{\boldsymbol{k}}^{mm'}\!)\hat{f}_{\boldsymbol{k}Bm\sigma}^{\dagger}\hat{f}_{\boldsymbol{k}Bm\sigma},
\end{align}
where we have introduced the quasiparticle weights $Z^\nu_{m\sigma}=|\langle \hat{O}_{i\nu m\sigma}^{\dagger}\rangle _{s}|^2$,
which act as renormalization factors for the hoppings, and $\tilde{\lambda}^\nu_{ m\sigma}=\lambda_{\nu m\sigma}-\lambda_{\nu m\sigma}^{0}$. In Eq.~\eqref{eq:H_f_AF}, $\varepsilon_{\boldsymbol{k}}^{mm'}=\underset{i}{\sum}t_{ij}^{A m B m'}e^{-i\boldsymbol{k}\cdot(\boldsymbol{R}_{i}-\boldsymbol{R}_{j})}$ while $\gamma_{\boldsymbol{k}}^{mm'}=\underset{i}{\sum}t_{ij}^{A m A m'}e^{-i\boldsymbol{k}\cdot(\boldsymbol{R}_{i}-\boldsymbol{R}_{j})}=\underset{i}{\sum}t_{ij}^{B m B m'}e^{-i\boldsymbol{k}\cdot(\boldsymbol{R}_{i}-\boldsymbol{R}_{j})}$ (note that the term $i=j$ is included here) and its expression does not depend on $j$ due to the translational invariance of the sublattices. 

Since the considered unit cell is able to host different types of symmetry breaking (e.g. a 2-site unit cell allows both N\'eel antiferromagnetism and two-site ferrimagnetism), the symmetry of the broken state can be enforced.
Indeed in the N\'eel phase all nearest-neighbor sites have the same magnetization, albeit in opposite directions. In the present framework this implies that the single-site slave-spin wave function for the $A$ and $B$ site, $|\phi_s^A\rangle$ and $|\phi_s^B\rangle$, will be identical under the exchange of their $\sigma=\uparrow$ part with the $\sigma=\downarrow$ part, and the same will hold for the couplings entering their single-site self-consistent Hamiltonian $H^A_s$ and $H^B_s$. This means that we can solve only one single-site slave-spin problem, calculate the average values entering $H_f$ (i.e. $\langle \hat{O}_{im\sigma}\rangle _{s}$ and $\langle \hat{S}_{im\sigma}^z\rangle _{s}$ for both $\sigma=\uparrow$ and $\sigma=\downarrow$) for the A site and use them exchanged for the B site.


We now specialize the above formalism to a more specific model (for an application to a full-fledged realistic Hamiltonian, see Section \ref{sec:BaCr2As2}). 

We here study the multi-orbital Hubbard model with M non-hybridizing degenerate orbitals ($t_{ij}^{mm'}=0$ for all $m\ne m'$, $\epsilon_m=0$). 
We restrict the hopping to nearest neighbors, i.e.  $t_{ij}^{mm}=t$ for $i \mathrm{ n.n. } j$, and zero otherwise. This entails that the only nonzero hopping is from one sublattice to the other, and thus $\gamma_{\boldsymbol{k}}^{mm'}=0$. Also $\varepsilon_{\boldsymbol{k}}^{mm'}\neq 0$ only for $m=m'$. 

In this case the fermionic Hamiltonian \eqref{eq:H_f_AF} decouples in orbital space and the quasiparticle bands are given by:
\begin{equation}
\Lambda_{\boldsymbol{k}}^\pm=\frac{\tilde\lambda^A+\tilde\lambda^B}{2}\pm \frac{1}{2}\sqrt{(\tilde\lambda^A-\tilde\lambda^B)^2+Z^{A}Z^{B}\varepsilon_{\boldsymbol{k}}^2}
\end{equation}
which are 2M times (for orbital and spin) degenerate (we have thus dropped the orbital and spin indices to lighten the notation). $\Delta=|\tilde\lambda^A-\tilde\lambda^B|$ is the gap opened in this band structure by the AF order.
 
In absence of the AF order ($\tilde\lambda^A=\tilde\lambda^B$ and $Z^{A}=Z^{B}$) one finds a unique (because the states are distinct in the NMBZ), spin- and orbital-degenerate renormalized band of which $\varepsilon_{\boldsymbol{k}}$ is the dispersion in the non-interacting case ($Z=1$, $\tilde \lambda=0$).

All the k-dependence of the problem enters through $\varepsilon_{\boldsymbol{k}}$ and this allows to define a density of states (DOS) $D(\varepsilon)=\frac{1}{2\mathcal{N}}\sum_{\boldsymbol{k}\in\mathrm{NMBZ}}\delta(\varepsilon-\varepsilon_k)$ (where $2\mathcal{N}$ is the total number of sites in the lattice). Both these quantities are determined by the geometry of the lattice.
We here choose to study a customary and particularly simple case of the Bethe lattice, which bares a semicircular DOS $D(\varepsilon)=\frac{2}{\pi D}\sqrt{1-\varepsilon^2/D^2}$ of half-width $D=2t$.

\subsection{One-band Hubbard model: benchmarks with slave-bosons and DMFT; phase separation}\label{sec:one-band}
\begin{figure}[h!]
\centering
\includegraphics[width=\columnwidth]{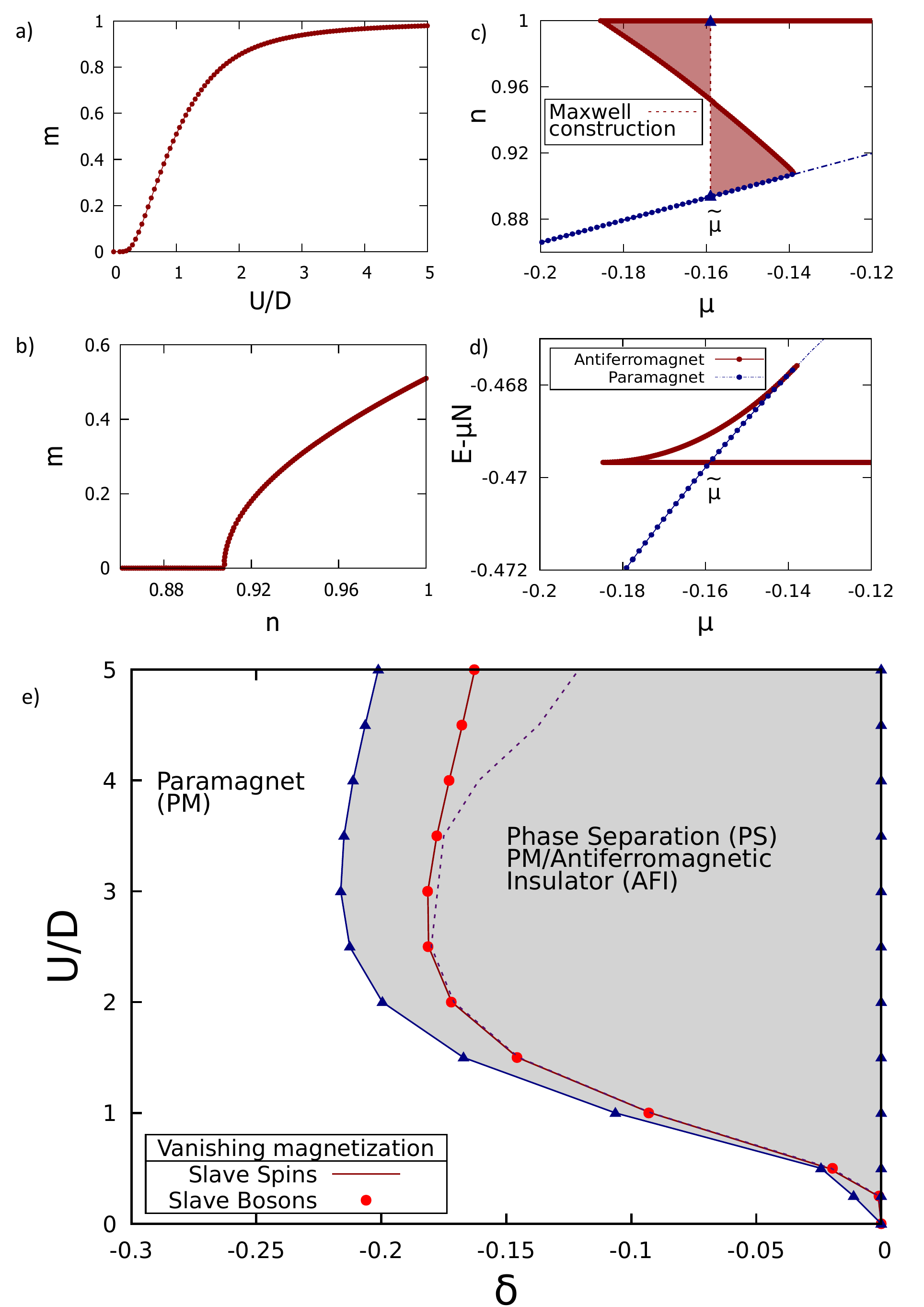}
\caption{\label{fig:1orb_united} Single-band Hubbard model on the Bethe lattice. a) Total staggered magnetization as a function of the
interaction strength at half-filling for the AF insulator. b) Magnetization as a function of density in the doped metallic phase, for the representative case of $U/D=1.0$. c) Density vs chemical potential at $U/D=1.0$, showing that in this model the AF metal is unstable (part of the curve with negative slope, i.e. negative compressibility). The dashed line marks the equal-area Maxwell construction, and connects the two stable solutions composing the phase-separated mixture associated with this first-order transition. The endpoints of the dashed line (blue triangles) indicate the coexisting phases (AFI at half filling and PM) of the mixture. The blue dot-dashed line shows the constrained paramagnetic solution obtained without allowing for the symmetry breaking. d) corresponding bow-shaped free energy $E -\mu N$ showing a crossing of the stable branches, which identifies the chemical potential of the first-order transition. e) Phase diagram. The red line marks the vanishing point of the magnetization (solid line: SSMF - points: SBMF), i.e. a diverging susceptibility of the paramagnetic metal to a staggered magnetic field. The dashed line is the spinodal where the compressibility $\kappa=dn/d\mu$ diverges. The light-grey zone indicates phase separation between the paramagnetic metal and the antiferromagnetic insulator.}
\end{figure}

As a first benchmark for our method, in Fig.~\ref{fig:1orb_united} we present results for the single band Hubbard model ($M=1$).

We show the on-site magnetization $m=n_{A\uparrow}-n_{A\downarrow}=n_{B\downarrow}-n_{B\uparrow}$ as a function of the interaction strength U at half-filling and as a function of doping $\delta=M-n$ (where $n=\langle\hat N\rangle/V$ is the total density) at fixed $U/D=1.0$. At half-filling an AF insulating (AFI) state is realized as soon as U is finite. The magnetization (Fig.~\ref{fig:1orb_united}a) increases quickly with $U$ and tends to saturate to a fully polarized state with $m=1$ as one expects on physical bases. This result is analogous to the one obtained within SBMF by Korbel et al. \cite{korbel2003antiferromagnetism} for a constant DOS.  
Upon doping, the AF state becomes metallic (AFM) and the magnetization decreases until vanishing (Fig.~\ref{fig:1orb_united}b). The line where this happens as a function of U is reported in red in the phase diagram (Fig.~\ref{fig:1orb_united}e), and signals the density at which the PM metallic solution develops an infinite susceptibility to a staggered magnetic field. 

Incidentally, all the calculated physical quantities, and thus the phase diagram, coincide exactly with those calculated within SBMF (see e.g. the same frontier as above, calculated with SBMF and reported as red points in the phase diagram). Indeed the present phase diagram is extremely similar to the one reported in the original work by Kotliar and Ruckenstein in Ref. \onlinecite{kotliar_ruckenstein} (albeit calculated there for a constant DOS).

Remarkably, the metallic AF phase is typically unstable in this model, i.e. it has negative compressibility $\kappa=dn/d\mu$ in most of the phase diagram. Its shape for a typical case ($U/D=1.0$) is shown in Fig.~\ref{fig:1orb_united}c. Including the flat plateau corresponding to the AF insulator and the PM metallic branch, the $n(\mu)$ curve has a sigmoidal shape, which implies the insurgence of a zone of phase separation in the phase diagram. This zone is determined by a Maxwell construction, where the chemical potential $\tilde{\mu}$ at which the phases coexist is determined by finding the vertical line $\mu = \tilde{\mu}$ which cuts the sigmoid into two closed equal-surface areas. The two endpoints of the $\mu=\tilde{\mu}$  line single out the two stable solutions that mix in the separated phase. The sigmoid in $n(\mu)$ corresponds to a bow shape of the free-energy $E-\mu N$ (where $E=\langle \hat H \rangle$ and $N=\langle \hat N \rangle$) of Eq.~\eqref{eqn:energy}, since $n(\mu)=-1/V\partial(E-\mu N)/\partial \mu$, and the chemical potential $\tilde{\mu}$ of the phase separated mixture corresponds to the point where the stable branches cross\cite{Callen_book}, as shown in Fig.~\ref{fig:1orb_united}d.
The average density imposed by the number of electrons physically present in the system constrains the proportions of the two components of the separated phase.

Since the two endpoints of the $\mu=\tilde{\mu}$ line are respectively on the PM metallic and on the AF insulating branches in all the cases we have analyzed in the single-band Hubbard model we can conclude that the system separates into these two phases and that the AF-PM transition is always first-order. In the phase diagram the two endpoints are marked by blue triangles and the corresponding zone of phase separation is grayed.

The two spinodal points where the compressibility is infinite and changes sign - which mark the limits the strictly unstable part of the homogeneous solution - are always inside the actual zone of phase separation. Between the spinodal and the endpoints of the Maxwell construction one finds metastable branches of the homogeneous solution (like the overheated of undercooled branches in the familiar example of the first-order liquid gas transition).
These points do not have to always coincide with the border of the AF metal as in Fig.~\ref{fig:1orb_united}c, other behaviours of $n(\mu)$ are encountered in this and other models, that shape different and more complex phase diagrams. Indeed at intermediate to large $U/D$ in the single-band model the $n(\mu)$ curve acquires rather a form similar to Fig~\ref{fig:E_mu-joined}b, with a positive compressibility at large doping, and thus the corresponding spinodal in the phase diagram (dashed line) departs from the one marking the vanishing magnetization.

\begin{figure}[h]
\centering
\includegraphics[width=\columnwidth]{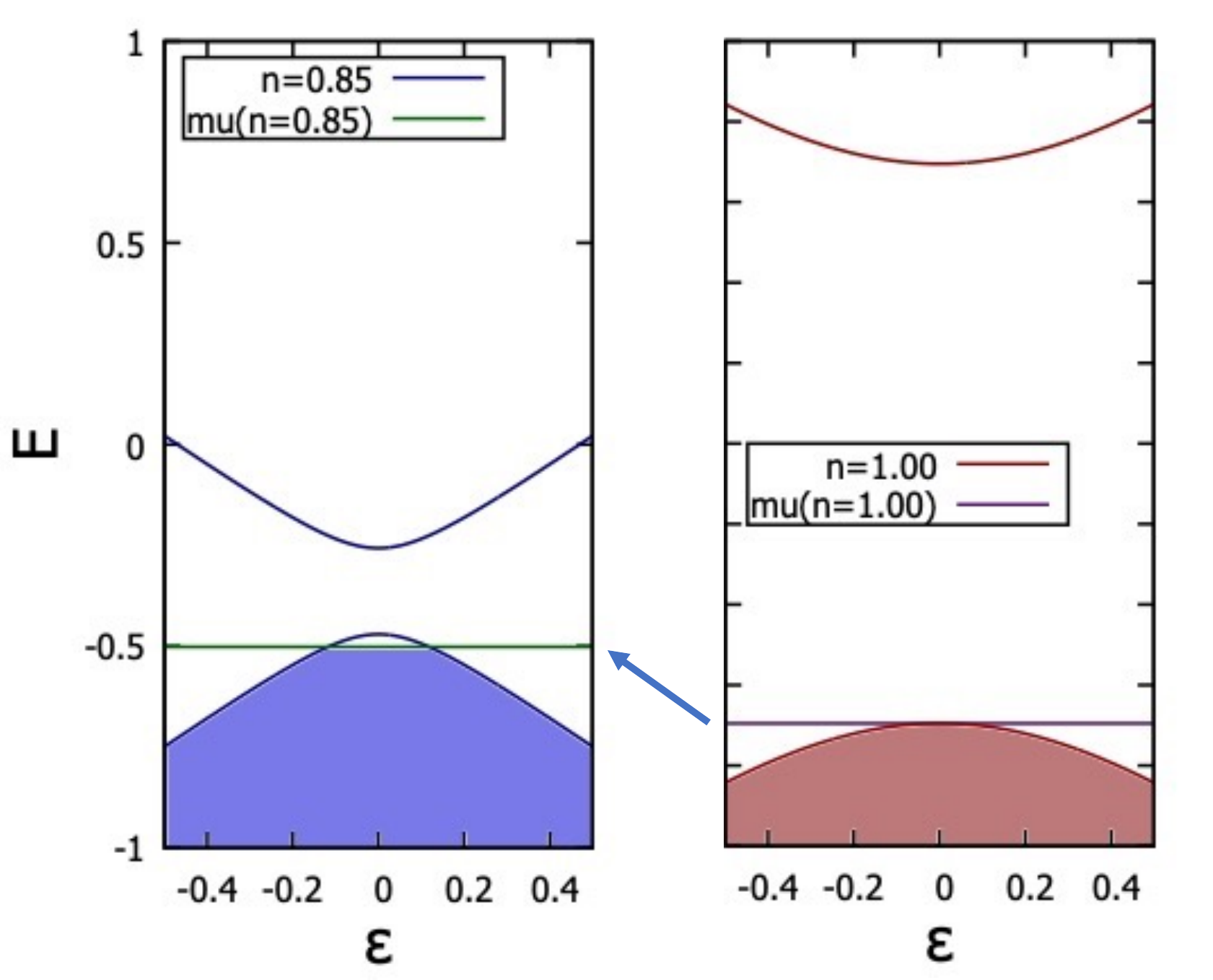}
\caption{\label{fig:gap_evol} Rationale behind the instability of the AF metal. Upon (hole-)doping the chemical potential enters the lower band while the gap shrinks, possibly resulting in a higher value of $\mu$ at lower density.}
\end{figure}

The counterintuitive behaviour of the chemical potential increasing with decreasing total population can be rationalized as follows (see Fig.~\ref{fig:gap_evol}, we here focus without loss of generality on hole doping). In the insulating AF phase at half filling, the chemical potential $\mu$ lies within the gap. Upon hole-doping $\mu$ enters the lower band. Meanwhile the gap starts closing, owing to a decreasing magnetization. There is thus a competition between these two effects, which can result in a chemical potential lying at a higher value for a lower density. The mechanism is robust, whereas the outcome depends quantitatively on the DOS and on the detailed doping dependence of the correlation-driven renormalization of bands' position and width. This has to be analyzed case by case.

Previously published works performed with SBMF and its rotationally-invariant extension also highlight the tendency of the AF state in the single-band Hubbard model towards phase separation and charge instabilities \cite{Camjayi-AF_Temp_Inst,Riegler_AF_neg_comp,Seufert-AF_2DHub_Charge-Inst}.
Moreover studies of this model within the more accurate DMFT framework validate further this physical picture. Indeed in Refs. \onlinecite{VanDongen_AF_PhaseSep,Zitzler_AF_PhaseSep} phase separation is found in the single-band Hubbard model for a hypercubic lattice, while for the Bethe lattice Ref. \onlinecite{Koch_Sangiovanni_Cluster_ED_AF} reports only a strong increase of the compressibility, which however remains finite. We have repeated this calculation and we instead find a negative compressibility in the same zone (see Appendix~\ref{app:SSMF_vs_DMFT}). This discrepancy is probably a question of numerical accuracy in a zone of the phase diagram where DMFT solutions are very hard to converge. In any case, albeit its quantitative impact has to be carefully assessed in each case, the tendency towards the phase separation instability is confirmed.

In this article we want to assess this tendency in the multi-orbital case, and measure the impact of Hund's coupling on it.

\subsection{Phase diagrams in the two- and three- band models: dependence on the Hund's coupling $J$}\label{subsec:2-3orb}

\begin{figure*}[t]
\begin{center}
\includegraphics[width=\textwidth]{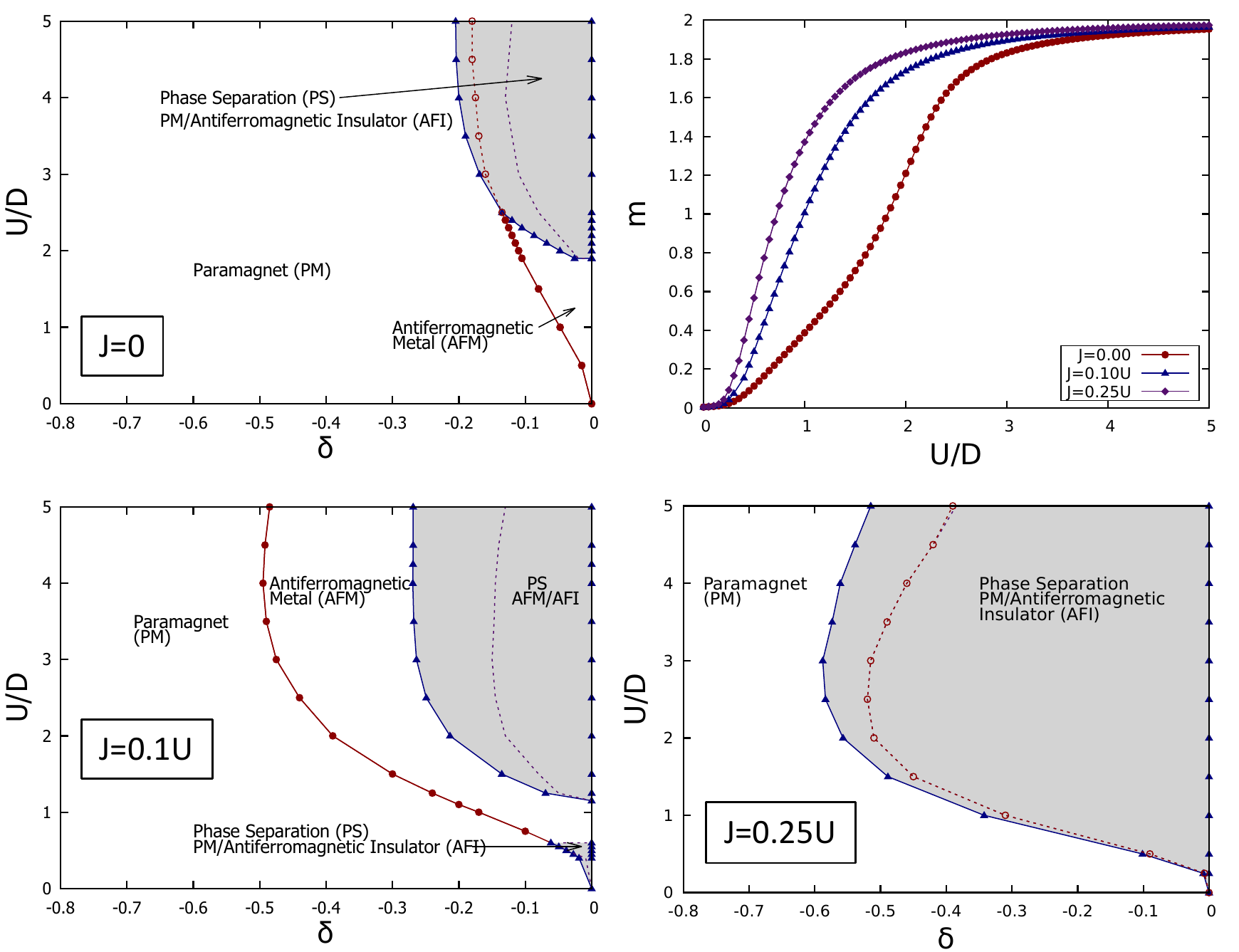}
\caption{\label{fig:2orb_PD} Two-orbital Hubbard model. On-site magnetization (top-right panel) for the AF insulator at half-filling and phase diagrams for three choices of Hund's coupling relative strength: $J=0$;  $J=0.1U$;  $J=0.25U$. The light-grey areas represent the phase separation zones; the red dots indicate a second order transition; the dashed line is the spinodal, where the compressibility diverges.}
\end{center}
\end{figure*}

Let's now consider the two-orbital case ($M=2$), for three values of Hund's coupling strength: $J=0$, $J/U=0.1$ and $J/U=0.25$.
The magnetization of the AF insulating state at half-filling is reported in Fig.~\ref{fig:2orb_PD} (top right). As a function of $U$ the magnetization increases, reaching saturation now obviously at $m\simeq2.0$. The results for the two orbital Hubbard model have the same general behaviour of the ones obtained by H. Hasegawa \cite{Hasegawa_SBMF_AF_2band} within the slave-boson framework for the simple cubic lattice. 
The effect of the Hund's coupling is to increase the magnetization, as one might expect, since it favors the high-spin configurations. The magnetization curve resembles the behaviour of the single-band case, rapidly reaching its saturation values with increasing $U$. In absence of Hund's coupling, the saturation is reached in a somewhat slower fashion. 

The phase diagrams as a function of doping are indeed also dependent on the value of Hund's coupling.
Our results for the three above-mentioned choices of $J/U$ are reported in Fig.~\ref{fig:2orb_PD}.

\begin{figure*}[t]
\begin{center}
\includegraphics[width=\textwidth]{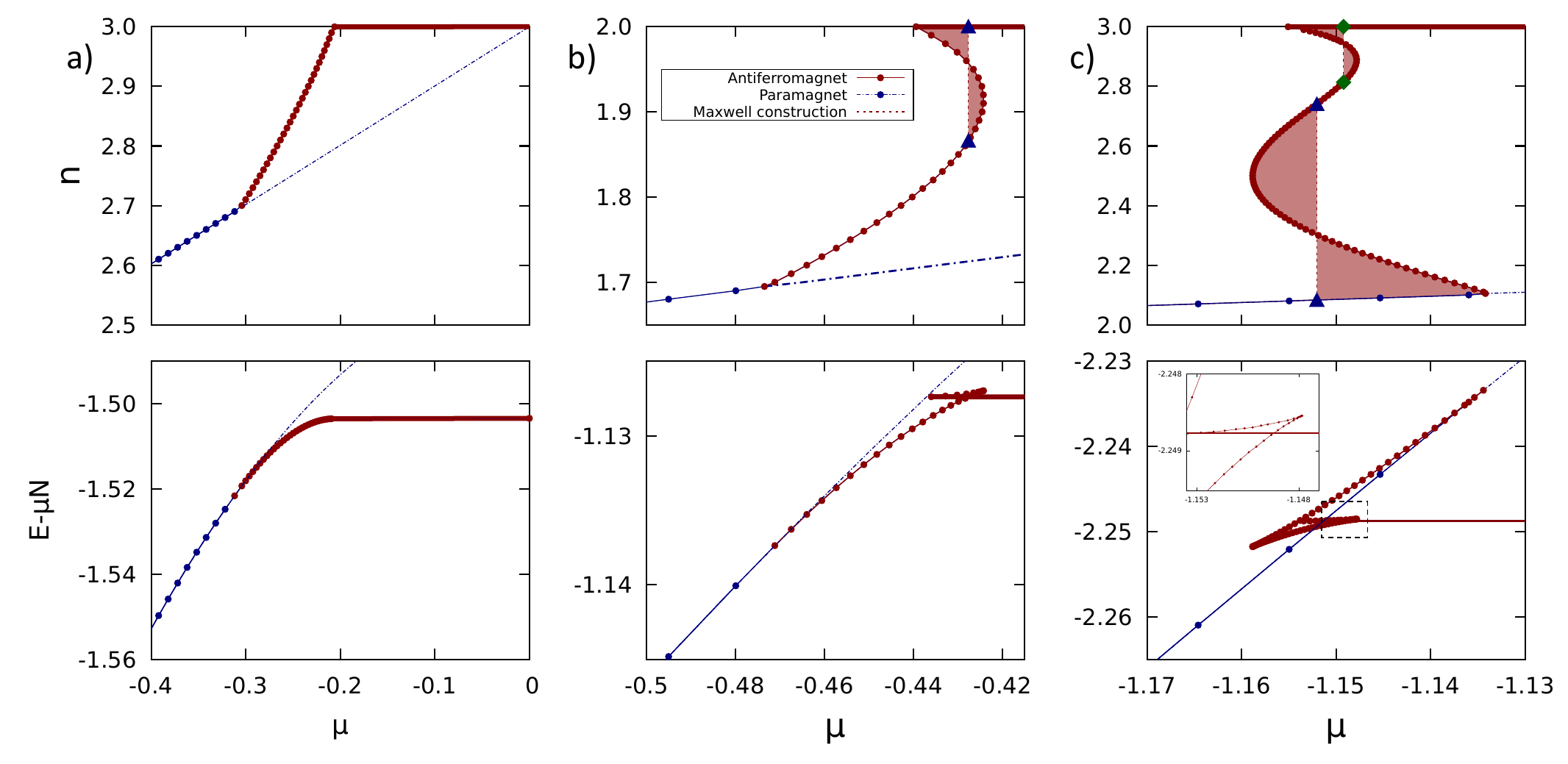}
\caption{\label{fig:E_mu-joined}Prototypical dependencies of the density (top panels) and corresponding free energy (bottom panels) as function of the chemical potential. Red and blue dots represent the antiferromagnetic and the
paramagnetic phase, respectively. The blue dot-dashed line shows the constrained paramagnetic solution obtained without allowing for the symmetry breaking. a) Gives a second-order transition between an AF metal and a PM metal (representative of $J=0$, $U/D\lesssim1.9$; $J=0.1U$, $0.60\lesssim U/D\lesssim1.15$ in the 2-orbital model and $J=0$, for all $U\lesssim 4.0$; $J=0.1U,\,$ $U/D\lesssim1.5$ in the 3-orbital model). b) Gives rise to a zone of phase separation between AF insulator and AF metal at low doping, and a 2nd-order transition at larger doping between an AF metal and a PM metal (representative of $J=0$, $1.9\lesssim U/D\lesssim2.5$; $J=0.1U$, $U/D\gtrsim1.15$ in the two-orbital model and $J=0.1U$, $U/D\gtrsim1.5$ in the three-orbital Hubbard model). If the sigmoid is much more pronounced the lower endpoint of the Maxwell construction can end up on the PM branch (representative of large $U/D$ in the 1-orbital model and in the multi-orbital cases at $J/U=0.25$, $J=0$, $U/D\gtrsim 2.5$ in the 2-orbital model $J=0$, $U/D\gtrsim 4.0$ in the 3-orbital model). As in Fig.~\ref{fig:1orb_united} the coexisting phases are marked by blue triangles. c) Gives two successive zones of phase separation (AFI-AFM, green diamonds) and (AFM-PM, blue triangles) with a zone of stable AFM in between (representative of  $J=0.25U$ and with $U/D\gtrsim1.66$ in the three orbital Hubbard model) as long as the upper red dashed line is on the right of the lower red dashed line. When they align, the two zones of phase separation touch and a coexistence of three phases is realized (a triple point, at $U/D~1.66$); for $U/D\lesssim1.66$ a unique Maxwell construction and phase separation zone (AFI-PM) remain.}
\end{center}
\end{figure*}

One main result is robust across all variations of models and Hund's coupling relative strength: the AF metal disappears into a PM metal at a certain doping away from half-filling. This doping value depends on U in a very similar fashion in all cases: it starts from zero at U=0, it reaches a maximum when U is a few times the bandwidth, before decreasing again for very high interaction strengths.
However the system does so in several different ways, which can all be characterized by the $n(\mu)$ dependence as we do in the following (see in Fig.~\ref{fig:E_mu-joined} for several examples, representative of the cases we analyzed), and which ultimately determine the order of the transition and the phases that one can observe.

For instance in the J=0 case at low $U/D\lesssim1.9$ the compressibility is positive (see $n(\mu)$ curve in Fig.~\ref{fig:E_mu-joined}a) and thus the AF metal is stable all the way to the doping where $m$ vanishes. The AF-PM transition is thus of the second order.
For $U/D\gtrsim1.9$ the $n(\mu)$ starts flexing into a sigmoidal form (as in the example Fig.~\ref{fig:E_mu-joined}b), as testified by the plotted spinodal point (dashed line in the phase diagram), and thus a phase separation occurs. Initially the Maxwell construction encompasses a small range of doping, so that for $1.9\lesssim U/D \lesssim 2.4$ the inhomogeneous state at low doping is a mixture of AF insulating and AF metallic phases. At larger doping the AF metal is stable until it becomes continuously a paramagnet.
For $U/D\gtrsim 2.4$ instead the sigmoid in $n(\mu)$ is so pronounced that the Maxwell construction connects directly the AF insulator and the PM metal, and the phase separation happens between these two phases.
\begin{figure*}[t]
\centering
\includegraphics[width=\textwidth]{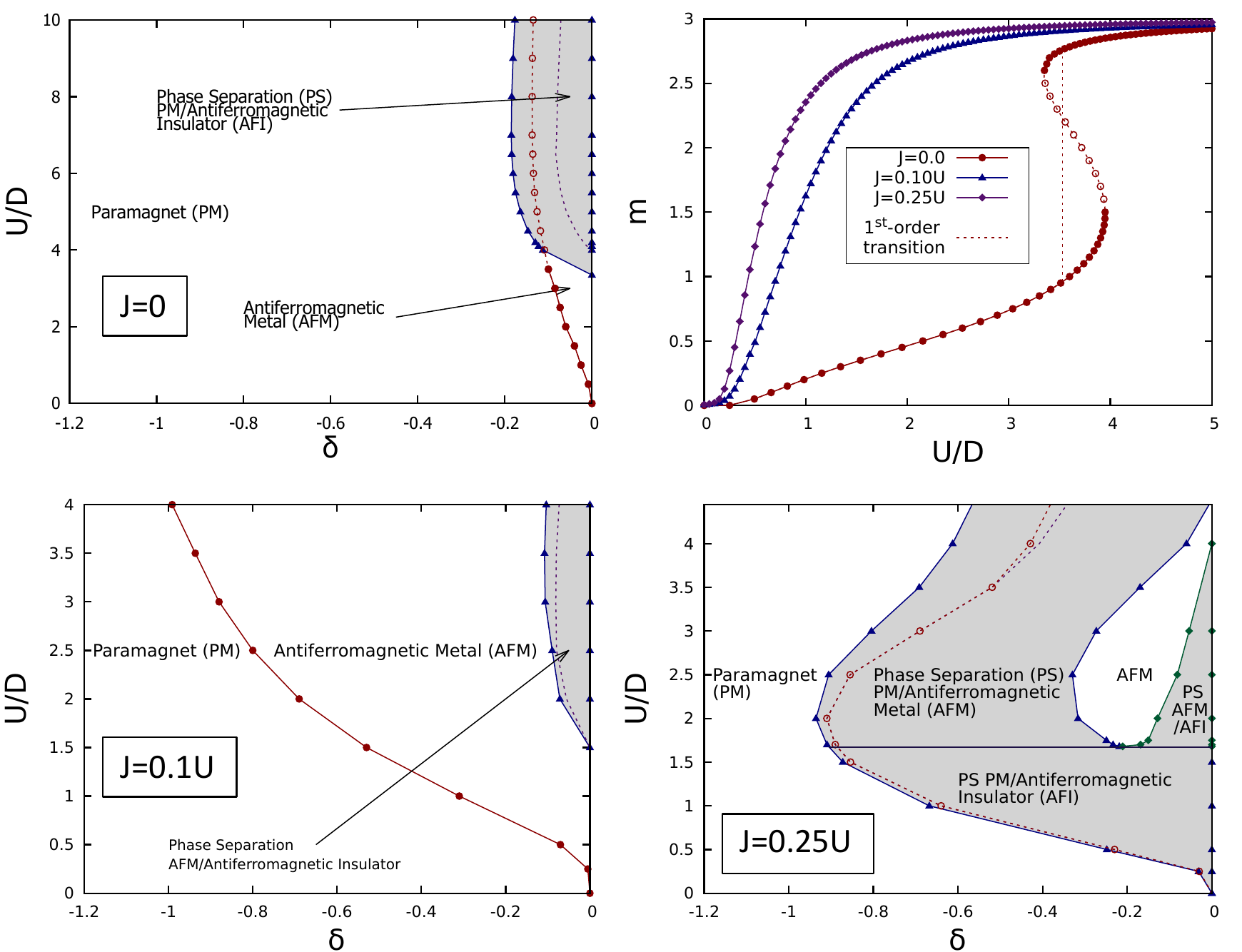}
\caption{\label{fig:3orb_PD}Three-orbital Hubbard model. On-site magnetization (top-right panel) for the AF insulator at half-filling and phase diagrams for three choices of Hund's coupling relative strength: $J=0$;  $J=0.1U$;  $J=0.25U$. The light-grey zones represent the phase separation zones; the red dots indicate a second order transition; the dashed lines is the spinodal, where the compressibility diverges; the solid black line indicates the region in the phase diagram equivalent to the triple point. }
\end{figure*}

The latter situation is never realized for the $J/U=0.1$, where the AF-PM transition is always second-order (besides a very small zone at $U/D\lesssim 0.5$ where $n(\mu)$ is of the type of Fig.~\ref{fig:1orb_united}c), because with $J$ the AF zone expands more than the phase separation one.

Instead for larger $J/U=0.25$ the phase separation zone catches up, becomes very wide in doping, and englobes entirely the AF metal. Indeed the $n(\mu)$ curve takes again a roughly straightened behaviour with negative slope exemplified in Fig.~\ref{fig:1orb_united}c, and the phase diagram becomes very similar to the one-band case.

On the methodological level, in all the cases we have checked, SSMF and SBMF results are identical, confirming also in the multi-orbital case that the two methods are equivalent (and equivalent to the GA) at $T=0$, as we pointed out in the one-band model.

More physically, the line of vanishing magnetization for $J/U=0.25$ in Fig.~\ref{fig:2orb_PD} (bottom-right) can be directly compared with the DMFT data shown in Fig.~3 of the work from Hoshino and Werner Ref.~\onlinecite{Hoshino_2band}, and show a good agreement, albeit the extent of the AFM phase is somewhat overestimated in our case. This can be ascribed to both finite-T effects in DMFT and to the more approximated nature of the SSMF compared to DMFT, which is likely to matter particularly at low magnetization (thus near the frontier) where dynamical fluctuations are more important. Furthermore, in degenerate models like the present one, the latter are likely to be enhanced with respect to more realistic models with e.g. crystal field splitting. Also SSMF are known to perform better when the number of orbitals gets larger\cite{demedici_Vietri}. These considerations are coherent with the result on the realistic 5-orbital benchmark case of BaCr$_2$As$_2$ we report in Section \ref{sec:BaCr2As2},  the agreement with DMFT is excellent.


Let's finally address the three-orbital ($M=3$) case.

The magnetization at half-filling as a function of $U/D$ is plotted in Fig.~\ref{fig:3orb_PD} (top-right panel). While at intermediate-to-large $J$ the physics is essentially the same as in the two-orbital model, a remarkable feature emerges in the $J=0$ case: a low-spin to high-spin transition at $U\approx4.0D$. Indeed a coexistence zone in $U$ is visible in the plot, where two stable solutions are found (connected by an unstable branch signalled by empty symbols).
This might seem surprising, in absence of the drive from Hund's coupling towards high-spin states. However the competition between the low- and the high-spin state has a natural rationale: while the high-spin phase maximizes the non-local AF exchange, the low-spin phase has a larger kinetic energy gain, due to the orbital fluctuations\cite{gunnarsson_multiorb,florens_multiorb,Chatzieleftheriou_RotSym}. 

This feature shows up in the phase diagram Fig.~\ref{fig:3orb_PD} (top-left panel), which is very similar to the two-orbital case (albeit shifted towards larger values of $U$). One difference is the fact that here the spinodal starts at larger $U$ (corresponding to the end of the coexistence zone at half-filling) compared to the beginning of the phase separated zone, which instead connects smoothly with the discontinuous jump between the two solutions at half-filling.

Analogously, the phase diagram for $J/U=0.1$ Fig.~\ref{fig:3orb_PD} (bottom-left panel) is very similar qualitatively to the two-orbital case. The main difference is quantitative: the AF metal extends to much larger dopings at the same value of $J/U$.

Also the phase diagram for $J/U=0.25$ (Fig.~\ref{fig:3orb_PD}, bottom-right panel) is very similar to the two-orbital one, in its main features, mainly a strong increase in the doping range where the phase separation is realized.

However its physics is actually richer. 
For small interaction strengths we find a phase separation between the antiferromagnetic insulator and the paramagnetic metal, as in the one- and two-orbital models.
For greater values of $U$, however, the $n(\mu)$ curve has the shape of a double sigmoid, as shown in Fig.~\ref{fig:E_mu-joined}c. Correspondingly the free energy has a double-bow shape and two crossing points, and hence two distinct Maxwell constructions are in order. The system then exhibits two distinct zones of phase separation if, as in the case reported in the figure, the construction at smaller density singles out a value of $\mu$ smaller than the construction at higher density (i.e. as long as the lower dashed line in the figure is on the left than the upper dashed line). The first phase separation zone, marked with the green diamonds in the phase diagram, is between the AF insulator at half-filling and an AF metal; the second one is between an AF and a paramagnetic metal.

This happens for values of $U/D\gtrsim 1.66$. Approaching this interaction strength from above the two crossing points come closer to one another, until they meet. Then the two Maxwell constructions merge into one, and a \emph{triple point} in the $U$-$\mu$ plane is realized, where the paramagnet, the antiferromagnetic metal and the half-filled antiferromagnetic insulator all coexist. In the phase diagram in the $U$-doping plane this is represented with a solid black line.

Again the vanishing magnetization line for the $J/U=0.25$ three-orbital model can be compared to the DMFT results of Hoshino and Werner\cite{Hoshino_SC} and the agreement is quite good, as in the two-orbital case, and actually slightly better. A somewhat overestimated width in doping of the AF-metal zone is obtained within SSMF. As already mentioned this overestimate is probably a shortcoming of SSMF,  due to the simplified treatment of dynamical fluctuations compared to DMFT, which is likely even more relevant in degenerate orbitals. This aspect improves substantially with the number of orbitals in the system, as the 5-orbital example of BaCr$_2$As$_2$ reported in Section  \ref{sec:BaCr2As2}, for which the agreement between SSMF and DMFT is excellent, testifies.

The fact that a large Hund's coupling can favor the phase separation instability can be rationalized within the same mechanism suggested in Section \ref{sec:one-band} and sketched in Fig.~\ref{fig:gap_evol}, since at larger J the magnetization is enhanced and so is the gap. The competition between the collapsing gap and the band population can more easily lean towards an inverted dependence of the chemical potential on the density, and thus towards a charge instability.

\section{Electronic correlations and itinerant vs localized magnetism}\label{sec:corr}

\begin{figure}[t]
\begin{center}
\includegraphics[width=\columnwidth]{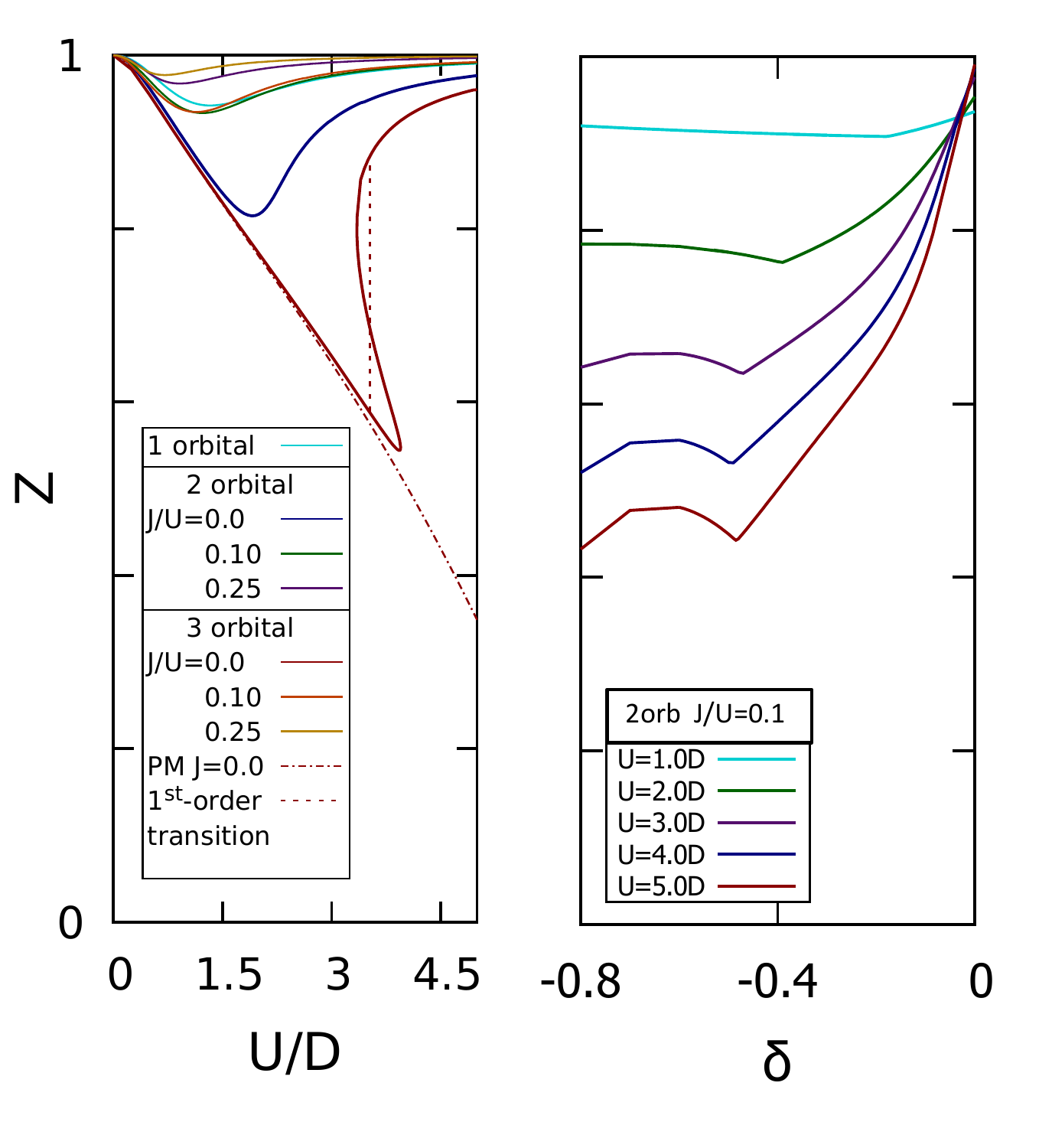}
\caption{\label{fig:Z} Quasiparticle weight Z as a function of $U/D$ at half-filling for all the models analyzed in this work (left panel), and as a function of doping for several values of $U/D$ for the two-orbital model at $J/U=0.1$. The kink in doping signals where the AF and PM metals connect and typically marks the lowest Z reached for each value of $U/D$. A larger number of orbitals and smaller value of J/U allow for more fluctuations between the local configurations and typically lead to more correlated states (e.g. the 3-orbital model at half-filling and $J=0$).}
\end{center}
\end{figure}

The SSMF captures band renormalization, which can be sizeable also in magnetic phases. However strongly polarized phases typically tend to minimize quantum fluctuations, thus reducing correlations.
Indeed it is known\cite{korbel2003antiferromagnetism} that in the AF phase of the one-band Hubbard model the quasiparticle weight Z is typically rather close to unity not only for small U/D, where correlations are obviously weak, but also at large U/D, because the magnetization saturates. 
We have here characterized the behavior of Z  in the M=2 and M=3 cases explored in this paper, and found that at half-filling - thanks to the quantum fluctuations allowed by the multi-orbital physics - Z is smaller the larger the number of orbitals, in absence of Hund's coupling. In the 3-orbital model in particular Z is very close to the PM value and nearly reaches 0.5 before undergoing the first-order low-to-high spin transition described in Sec.~\ref{subsec:2-3orb}, where Z markedly increases back towards unity. This behaviour is visible in the Z(U) curves reported in Fig.~\ref{fig:Z} (left panel). A finite $J$, however, it tends to quench the quantum fluctuations and to bring the magnetization quickly towards saturation, entailing very small correlations strengths.

Moreover, the quasiparticle weight is typically the highest at half-filling in the AF phase, due to the peak in magnetization (as expected in our particle-hole symmetric models). We studied its behaviour as a function of doping and found that it typically diminishes monotonously with it, until reaching the PM, where it starts increasing again. The typical behaviour for different values of $U/D$ is reported in Fig.~\ref{fig:Z} (right panel) for the 2-orbital model at $J/U=0.1$. The same curves for the three-orbital model are similar.

\begin{figure}[h]
\begin{center}
\includegraphics[width=\columnwidth]{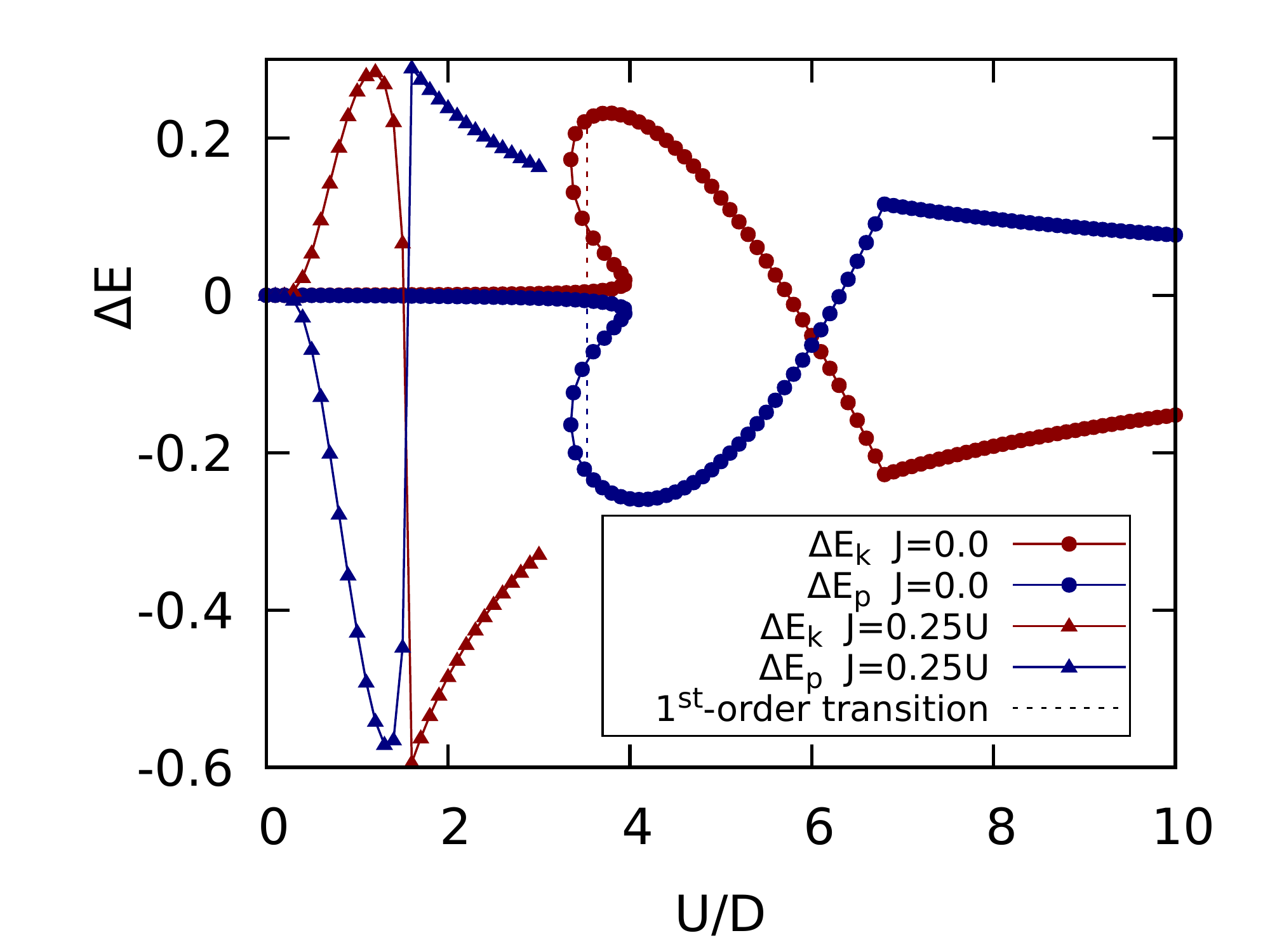}
\caption{\label{fig:Kin_Pot_Energies}Kinetic ($\Delta E_k=E_k^{AF}-E_k^{PM}$) and potential ($\Delta E_p=E_p^{AF}-E_p^{PM}$) energy differences between AF and PM phase. The total energy difference $\Delta E_{tot}=\Delta E_k+\Delta E_p < 0$, hence AF is always the stable phase. Data are plotted for $J=0$ (dots) and $J=0.25U$ (triangles) in the three-orbital Hubbard model. The dashed line marks the first-order low-high spin transition of the $J=0$ case, as reported in Fig.~\ref{fig:3orb_PD}.}
\end{center}
\end{figure}

Besides the band renormalization another highly nontrivial effect of the dynamical correlations is the evolution from itinerant to localized magnetism.
This is hardly visible in the mean-field solution for the ground state\cite{Sangiovanni-Static_vs_Dynamic-AF}, but a comparison with the PM solution clarifies the mechanism stabilizing the AF solution.
Indeed Taranto et al. report (Fig. 3 in Ref.~\onlinecite{Taranto-Optics_AF}) the difference in kinetic and potential energies of the two phases as a function of U/D in the single-band model at half-filling within DMFT. The AF, which is the stable solution at all interaction strengths, has however a lower potential energy than the parmagnet at small U/D, which then overcompensates an opposite raise in the kinetic energy and is thus the stabilizing factor for the AF. Conversely, at large U/D the PM has a lower potential energy, and it is the kinetic energy that wins over it and stabilizes the magnetic phase. Our method perfectly reproduces this behavior in the single-band model (not shown) and confirms this analysis in all the multi-orbital models (the curves for the M=2 and M=3 case sharing most of the qualitative features, we show only the latter), as we illustrate in Fig.~\ref{fig:Kin_Pot_Energies}. 

Furthermore, it is interesting to point out some model-dependent features. Indeed the figure reports these energy differences $(\Delta E_k=E_k^{AF}-E_k^{PM}$ and $\Delta E_p=E_p^{AF}-E_p^{PM})$ for the three-orbital model at zero and finite Hund's coupling $J$.
At strong coupling in the $J=0$ case (on the right in the figure) one finds the aforementioned energy balance $\Delta E_p>0$ rather insensitive to changes in U/D, due to the saturated AF and the Mott insulating PM showing the same behavior. The Mott transition in the PM solution, which at zero temperature is continuous if $J=0$\cite{Klejnberg_Spalek_Hund,florens_multiorb,Lechermann_RISB,demedici_Vietri} appears as a "kink". Thus, interestingly, the expected crossing of $\Delta E_k$ and $\Delta E_p$ when going towards the weak coupling opposite energetic balance happens well into the metallic phase. It is also interesting to note that the low/high spin transition we previously found in this model (corresponding to the dashed line cutting the heart-shaped feature) happens well within the weak-coupling regime $\Delta E_p<0$ and it is associated to a large potential energy difference between these two solutions.
For the large $J/U$ case one notices instead (besides the absence of low-high spin transition indeed, and the already discussed fact that all the action happens at much lower values of $U/D$) that, being the Mott transition in the PM phase of the first-order in this model\cite{Ono_multiorb_linearizedDMFT,Klejnberg_Spalek_Hund,Lechermann_RISB,demedici_Vietri,Chatzieleftheriou_Monster}, the continuous crossing between the weak- and strong-coupling regimes is cut away and happens as a sudden jump at the transition point.


\section{\emph{Ab-initio} description of G-type Antiferromagnetism in $BaCr_2As_2$.}\label{sec:BaCr2As2}
\begin{figure}[t]
\begin{center}
\includegraphics[width=\columnwidth]{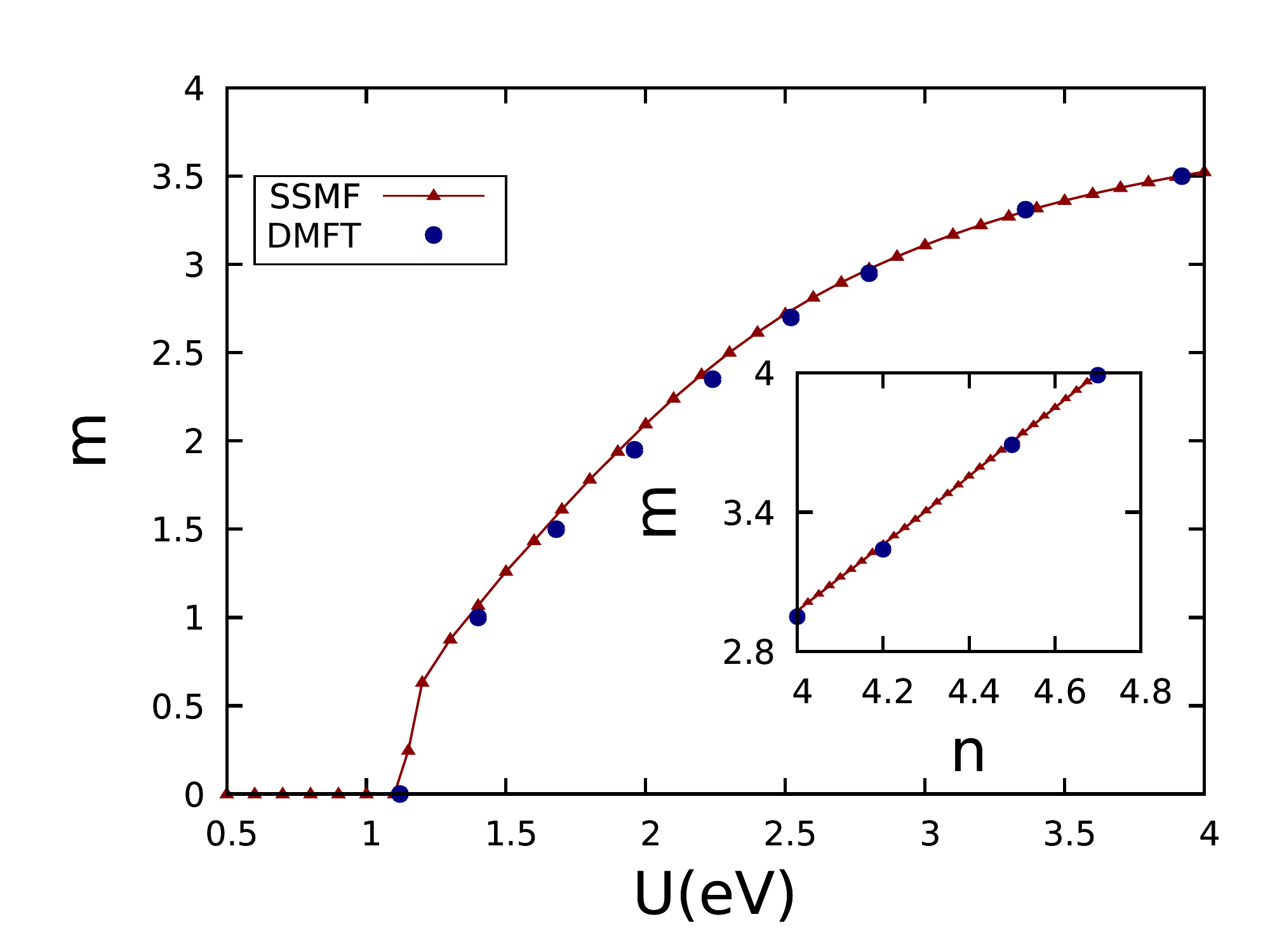}
\caption{\label{fig:BaCr2As2}Comparison between DFT+SSMF and DFT+DMFT for the G-type AF metallic phase of BaCr$_2$As$_2$. DMFT data from~\cite{Edelmann_Chromium_analogs}. In both methods the same parameters are used: $J=0.153U$; $n=4$ for the $U-$scan, and $U=2.8 eV$ for doping scan (inset).}
\end{center}
\end{figure}

As a final result we report a test of our method in the context of realistic \emph{ab-initio} simulations. We choose the case of BaCr$_2$As$_2$, for which the G-type AF phase was explored in Ref.~\cite{Edelmann_Chromium_analogs}\footnote{All the details about DFT calculations and the tight-binding model derived from it are reported in Sec II of Ref.~\cite{Edelmann_Chromium_analogs}}. We study the magnetization both as a function of the interaction strength and of doping. Our results are shown in Fig.~\ref{fig:BaCr2As2}, where we report the comparison of SSMF with DMFT, which is excellent. 
This agreement is particularly good both because, as already stated, the SSMF is known to perform better for larger number of orbitals\cite{Fanfarillo_Hund,demedici_Vietri}, and probably because of the reduced amount of quantum fluctuations in an ordered phase, compared to the more challenging case of the paramagnetic correlated metals. Still, the prospect for the use of the present method in realistic DFT-based quantitative modeling of correlated magnetic phases is very promising and further application to the 122 iron superconductor family are foreseen. 

A rough estimate of the different computational effort required by two methods for this specific example is the following: a complete iteration in SSMF takes O(seconds) on 1 processor, while in DMFT it takes O(hours) on O(1k) processors\cite{Sangiovanni-w2dynamics}\footnote{Despite SSMF is undeniably computationally less expensive, DMFT is still more accurate and it allows for a wider study of physical quantities. Also, the agreement in the realistic case presented here is excellent regarding the static observable that is the staggered magnetization, but other (typically dynamical) physical observables will certainly differ more, in particular in regimes characterized by strong quantum fluctuations.}. 

\section{Conclusions}
In summary, we have provided a variational derivation of the Z$_2$-Slave-spin mean field and shown that it brings an additional term containing an effective orbital energy shift to the known equations, and make them coincide with the U(1)-SSMF if the self-consistent Hamiltonian is real. 

We have shown that this method is capable of modeling spontaneous symmetry-broken phases, and provides identical results to Kotliar-Ruckenstein Slave-bosons mean-field and its multi-orbital generalizations, and hence to the Gutzwiller approximation, for the ground state of the system (thus at equilibrium at T=0).

We have then studied the N\'eel antiferromagnetic phases in the single-orbital, 2-orbital and 3-orbital Hubbard model. In the former we have highlighted a general tendency towards electronic phase-separation between an AF insulator and a PM metal in the doped phase near half-filling. 

In the multi-orbital cases we have studied the U-doping phase diagrams as a function of the Hund's coupling strength ($J=0$, $J/U=0.1$, and $J/U=0.25$) and shown that the phase separation can also happen between an AF insulator and AF metal and between an AF metal and a PM metal. The AF-PM transition can be first or second-order depending on the case at hand. A general trend that we have highlighted is that large Hund's coupling widens the AF range in doping and typically enhances the tendency towards phase-separation. We have provided a general rationale for such a tendency and its enhancement with increasing $J/U$: a competition between the depopulation of the renormalized band structure below the gap for hole-doping (or of the population of the renormalized band structure above the gap for electron doping) and of the closing of the gap itself, can cause the chemical potential to lie higher at smaller density in some range of parameters, thus causing negative compressibility and the related instability.

\acknowledgements The authors are supported by the European Commission through the ERC-CoG2016, StrongCoPhy4Energy, GA No724177. MC and LdM thank G. Sangiovanni for useful exchanges and discussions. 


\appendix

\section{Benchmarks with DMFT on models.}\label{app:SSMF_vs_DMFT}

The accuracy of the method we have presented in this work is naturally benchmarked relatively to existing slave-particle mean-field methods, like SBMF (and the equivalent GA). We have shown that all these methods produce strictly equivalent results for the ground state. 
A comparison in sample cases with the more accurate and well established DMFT is indeed also in order, to assess the overall reliability of the SSMF method for broken symmetry cases.
We have already cited Refs.~\onlinecite{Hoshino_2band,Hoshino_SC} for benchmarks in the two-orbital and three-orbital model cases. For a comparison in the context of ab-initio calculations we have shown the excellent agreement with the DMFT data of Ref.~\onlinecite{Edelmann_Chromium_analogs}.

\begin{figure}[t]
\begin{center}
\includegraphics[width=\columnwidth]{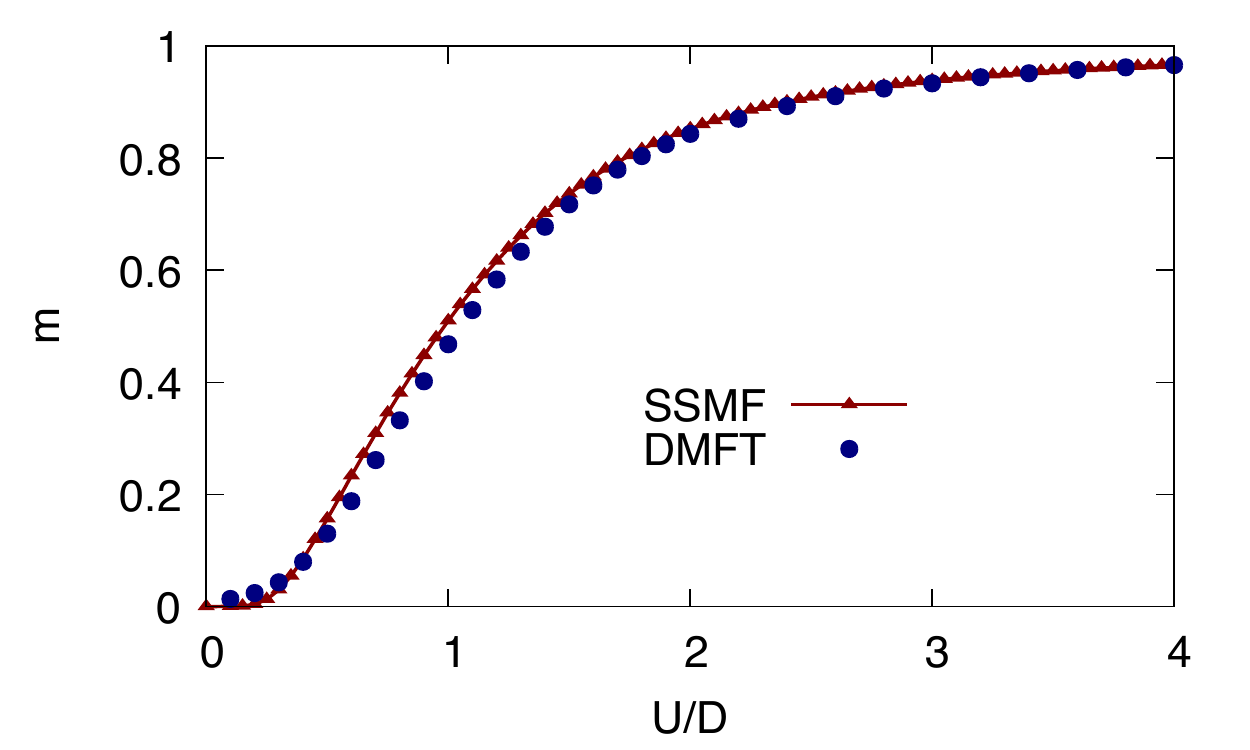}
\caption{\label{fig:DMFT_vs_SSMF_1band} Staggered magnetization in the one-band Hubbard model at half-filling: SSMF vs DMFT.}
\end{center}
\end{figure}

Here we show explicitly the comparison with DMFT in the single-band model. As a solver we use Exact Diagonalization\cite{Caffarel_Krauth} at T=0 with a total $N_s=6$ orbitals (5 sites in the bath and 1 on the impurity).
In Fig.~\ref{fig:DMFT_vs_SSMF_1band} we report the staggered magnetization vs $U/D$ in the one-band Hubbard model at half-filling. The agreement between the result of the two methods is very good. 
In fact, the agreement is here much better than in the paramagnetic phase, most likely because of the reduced quantum fluctuations in the ordered phase.

\begin{figure}[t]
\begin{center}
\includegraphics[width=\columnwidth]{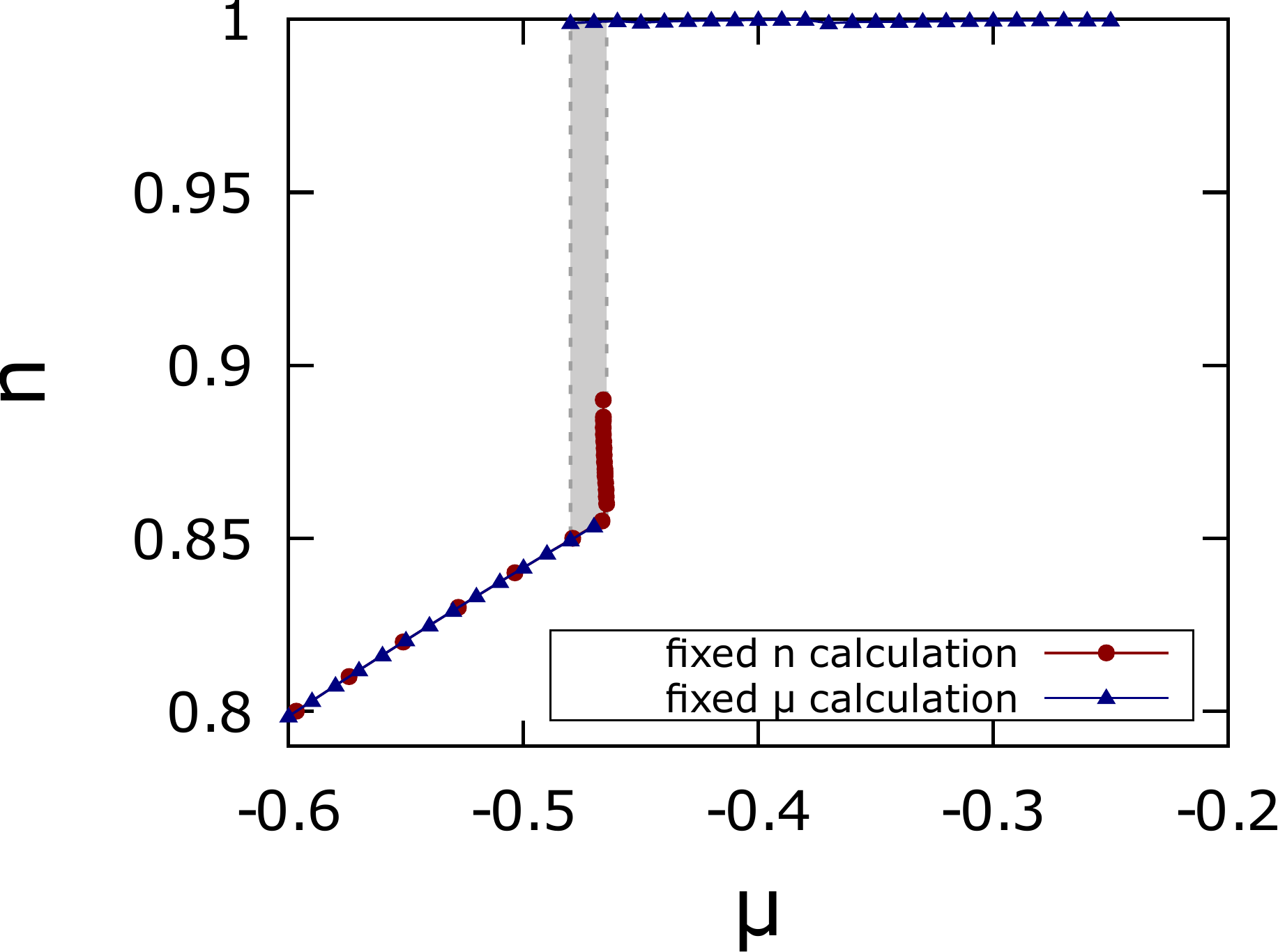}
\caption{\label{fig:DMFT_vs_SSMF_1band_doped} DMFT in the one-band Hubbard model: AF insulator - PM metal transition in doping. The range of chemical potential for which the AFI and PM coexist implies a first-order transition between them, and an unstable branch with negative compressibility joining them, which we could converge for a range of doping.}
\end{center}
\end{figure}

We also benchmark against DMFT the finding of negative compressibility implying the phase separation in the doped AF metal near half-filling, typically for the single-band model. In Fig.~\ref{fig:DMFT_vs_SSMF_1band_doped} we report our DMFT calculation for the Bethe lattice at $T=0$ and $U/D=2.0$. It is easy to see that there is a range of chemical potential for which the AF insulator (at density n=1) coexists with the PM metal at finite doping. It is very hard to stabilize the unstable branch connecting these two stable solutions and we succeded for a range of densities (by continuously adjusting $\mu$ in order to reach a given density\cite{Chatzieleftheriou_Monster}), confirming the negative compressibility found by SSMF (and found by DMFT in the hypercubic lattice in Refs. \cite{VanDongen_AF_PhaseSep,Zitzler_AF_PhaseSep}).

We finally underline that in the paramagnetic phases the quantitative agreement with DMFT improves substantially for systems with a large number of orbitals\cite{Fanfarillo_Hund,demedici_Vietri}. Indeed realistic investigations of iron-based superconductors (typically modeled with 5-orbital models) are performed routinarily with good results (albeit typically an adjustment of the $J/U$ value is required)\cite{YuSi_LDA-SlaveSpins_LaFeAsO,demedici_OSM_FeSC,demedici_Vietri,Yi_Universal_OSM_Chalcogenides}.
It is in line with this indication that the results obtained by SSMF and DMFT for the realistic 5-orbital case of the AF phase in BaCr$_2$As$_2$ and reported in Fig.\ref{fig:BaCr2As2} agree even better than in the single-band model case.

\bibliography{../Bib/publdm,../Bib/bibldm,../Bib/FeSC}

\end{document}